\documentclass[prd,eqsecnum,twocolumn,amsfonts,amssymb,amsmath, nofootinbib]{revtex4}
\pdfoutput=1

\usepackage{url}
\usepackage{hyperref} 
\hypersetup{
	colorlinks=true,
	citecolor=,
	linkcolor=,
	filecolor=,      
	urlcolor=blue,
}
\usepackage{graphicx}
\usepackage{caption}
\usepackage{subcaption}
\usepackage{bm}
    \usepackage[x11names]{xcolor}
\usepackage{framed}
\colorlet{shadecolor}{blue!10}

\newcommand{\beq}{\begin{equation}}
\newcommand{\eeq}{\end{equation}}
\newcommand{\beqs}{\begin{eqnarray}}
\newcommand{\eeqs}{\end{eqnarray}}
\newcommand{\lsim}{\mathrel{\raisebox{-
.6ex}{$\stackrel{\textstyle<}{\sim}$}}}

\begin{document}

\title{Cross Section Calculations in Theories of Self-Interacting Dark Matter}

\author{Sudhakantha Girmohanta and Robert Shrock}

\affiliation{\ C. N. Yang Institute for Theoretical Physics and
Department of Physics and Astronomy, \\
Stony Brook University, Stony Brook, New York 11794, USA }

\begin{abstract}

  We study an asymmetric dark matter model with self-interacting dark
  matter consisting of a Dirac fermion $\chi$ coupled to a scalar or
  vector mediator, such that the reaction $\chi + \chi \to \chi +
  \chi$ is well described by perturbation theory.  We compute the
  scattering cross section $\sigma$, the transfer cross section
  $\sigma_T$, and the viscosity cross section $\sigma_V$ for this
  reaction.  As one part of our study, we give analytic and numerical
  comparisons of results obtained with the inclusion of both
  $t$-channel and $u$-channel exchanges and results obtained in an
  approximation that has often been used in the literature that
  includes only the $t$-channel contribution.  The velocity dependences
  of these cross sections are studied in detail and shown to be in
  accord with observational data. 
  
\end{abstract}

\maketitle


\section{Introduction}
\label{intro_section}

There is compelling evidence for dark matter (DM), comprising about 85
\% of the matter in the universe.  Cold dark matter (CDM) has been
shown to account for the observed properties of large-scale structure
on distance scales larger than $\sim 10$ Mpc~\cite{primack, swo, nfw,
  nfw2, kravtsov_primack, moore, fsw2020}~\footnote{ {See, e.g.,
    Particle Data Group, Review of Particle Properties online at
    \href{http://pdg.lbl.gov}{http://pdg.lbl.gov}} and L. Baudis and
  S. Profumo,
  \href{chrome-extension://efaidnbmnnnibpcajpcglclefindmkaj/https://pdg.lbl.gov/2020/reviews/rpp2020-rev-dark-matter.pdf}{Dark
    Matter Minireview} at this website.\label{pdg_dm}} ~\footnote{
  Specifically, defining $\Omega_i \equiv \rho_i/\rho_c$, where
  $\rho_c = 3H_0^2/(8\pi G)$, with $H_0$ the current Hubble constant,
  $G$ the Newton gravitational constant, and $\rho_i$ the mass density
  of a constituent $i$, current cosmological observations yield the
  results $\Omega_m = 0.315(7)$ for the matter density,
  $\Omega_{DM}=0.265(7)$ for the dark matter density, and $\Omega_b =
  0.0493(6)$ for the baryon matter density (see
  \href{http://pdg.lbl.gov}{Particle Data Group
    online}). \label{dmv}}~(reviews include \cite{jkg,
  binney_tremaine, bhs, strigari, lisanti_review,dm_history}.)  Some
possible problems with fitting observational data on length scales of
$\sim 1-10$ kpc were noticed with early CDM simulations that lacked
baryon feedback \cite{spergel_steinhardt,boylan_kolchin,
  boylan_kolchin2}.  These included the prediction of greater density
in the central region of galaxies than was observed (the core-cusp
problem), a greater number of dwarf satellite galaxies than were seen
(the missing satellite problem), and the so-called ``too big to fail''
problem pertaining to star formation in dwarf satellite galaxies.
This led to the consideration of models in which dark matter particles
have significant self interactions. The extension of cold dark matter
$N$-body simulations to include baryon feedback can ameliorate these
problems with pure CDM simulations \cite{springel, springel2,
  scannapieco, fire2015, fire2016, nfw_apostle,
  bullock_boylan_kolchin, peter, fbbb, chua_vogelsberger_hernquist,
  springel2019, fire2}. Nevertheless, cosmological models with
self-interacting dark matter (SIDM) are of considerable interest in
their own right and have been the subject of intensive study
\cite{spergel_steinhardt}, \cite{dssw,kusenko_steinhardt,
mirror1,
bullet_cluster_constraints,
afsw,
feng_kaplinghat_yu,
buckley_fox,
koda,
vogelsberger_zavala_loeb,
kouvaris2012,
tulin_yu_zurek2013prl,
tulin_yu_zurek2013,
zurek_adm_review,
wise,
kusenko,
frandsen_sarkar,
elbert_bullock_peter,
kaplinghat_tulin_yu_prl2016,
slatyer_sommerfeld,
valli_reassessment,
kkpy,
battaglieri,
robertson2017,
robertson2018,
tulin_yu_review,
valli_yu,
lesgourgues_brinckmann_sidm,
brinckmann_vogelsberger,
vogelsberger_zavala_slatyer,
robertson2019,
rkky,
essig_sidm,
wechsler,
apr,
hayashi,
kaplinghat2020,
bondarenko,
essig_loverde,
tulin_velocity,
tulin_etal2021,
ebisu,
fischer2022a,
lisanti2022,
bullock_kaplinghat_valli2022,
eckert2022,
fischer2022}.  Other candidates for
dark matter, such as primordial black holes \cite{carr_pbh}, mirror
dark matter \cite{mirror1,mirror2,mirror3}, warm dark matter
\cite{bot,aks,wdm,frenk_wdm}, ultralight (pseudo)scalar dark matter
\cite{fuzzy_dm,axion_dm}, and dark matter in the context of
extra-dimensional models \cite{nuled,dmled} have also been studied but
will not be discussed here.

A general estimate shows what size the cross section for
scattering of dark matter particles, denoted generically as $\sigma$,
should be in order to alleviate problems with CDM simulations lacking
baryon feedback.  It is necessary that there should be one or more
DM-DM scatterings over the age of the universe. The rate of DM-DM
scatterings is given by
\beq \Gamma = \bigg ( \frac{\sigma}{m_{\rm DM}} \bigg ) \, v_{\rm rel} \,
\rho_{_{\rm DM}} \ ,
\label{gamma_sidm}
\eeq
where $m_{\rm DM}$ denotes the mass of the DM particle. Numerically, this is 
\beqs
\Gamma &=& 0.1 \, {\rm Gyr}^{-1} \,
\bigg ( \frac{\sigma/m_{\rm DM}}{1 \ {\rm cm}^2/{\rm g}} \bigg ) \,
\bigg ( \frac{v_{\rm rel}}{ 50 \ {\rm km/s}} \bigg ) \,
\bigg ( \frac{\rho_{_{\rm DM}}}{0.1 \ M_\odot/{\rm pc}^3 } \bigg ) . \cr\cr
&&
\label{gamma_sidm_numerical}
\eeqs
An important property of cross sections of self-interacting dark
matter particles, inferred from fits to observational data, is that
they should decrease as a function of the relative velocities $v_{\rm
  rel}$ of these DM particles.  Quantitatively, fits to galactic data
on the scale of $\sim 1-10$ kpc, with velocities $v_{\rm rel} \sim
50-200$ km/s, yield values $\sigma/m_{DM} \sim 1$ cm$^2$/g, while fits
to observations of galaxy clusters on distance scales of several Mpc
and $v_{\rm rel} \sim O(10^3)$ km/s generally yield smaller values of
$\sigma/m_{DM} \sim 0.1$ cm$^2$/g (note the conversion relation 1
cm$^2$/g = 1.8 barn/GeV).

In this paper we consider SIDM models in which the dark matter is
comprised of a spin-1/2 Dirac fermion $\chi$, interacting with a
mediator, generically denoted $\xi$.  Both the DM fermion and the
mediator are taken to be singlets under the Standard Model (SM). We
study two versions of this model, namely one in which the mediator
field is a real scalar, $\phi$, and another in which the mediator is a
vector field, $\xi=V$.  In both versions, we work in the context of an
asymmetric dark matter (ADM) theory (for a review, see, e.g.,
\cite{zurek_adm_review}).  Thus, by the time at which
large-scale structure formation begins, a net asymmetry has built up
in the number density of $\chi$ and $\bar\chi$ particles. By convention,
we take this asymmetry to be such that the number density of $\chi$ particles
is dominant over that of $\bar\chi$ particles. We assume
parameter values such that the lowest-order perturbative calculation
of the cross section gives a reliable description of the physics, so
we do not need to deal with nonperturbative effects and bound states
of dark matter particles. We compute the scattering cross section
$\sigma$, the transfer cross section $\sigma_T$, and the viscosity
cross section $\sigma_V$ for this reaction.  As one part of our study,
we give analytic and numerical comparisons of results obtained with
the inclusion of both $t$-channel and $u$-channel exchanges and
results obtained in an approximation that has often been used in the
literature that includes only the $t$-channel
contribution.  Our new results provide improved accuracy for fitting
models with self-interacting dark matter to observational data.

In the version of our SIDM model with a real scalar mediator $\xi =
\phi$, we take the interaction between $\chi$ and $\phi$ to be of
Yukawa form, as described by the interaction Lagrangian
\beq
    {\cal L}_y = y_\chi [\bar\chi \chi]\phi \ .
\label{chi_chi_phi}
\eeq
In the second version, the DM fermion $\chi$ is assumed to be charged
under a U(1)$_V$ gauge symmetry with gauge field $V$ and gauge
coupling $g$.  Since only the product of the U(1)$_V$ charge of $\chi$ times
$g$ occurs in the covariant derivative in this theory, we may, without
loss of generality, take this charge to be unity and denote the
product as $g_\chi$.  The corresponding interaction Lagrangian is
\beq
{\cal L}_{\bar\chi\chi V} = g_\chi [\bar\chi \gamma_\mu \chi]V^\mu \ . 
\label{chi_chi_v}
\eeq
A Higgs-type mechanism is assumed to break the U(1)$_V$ symmetry and
give a mass $m_V$ to the gauge field $V$.  For compact notation, we
use the same symbol, $\alpha_\chi$, to denote $y_\chi^2/(4\pi)$ for the
case of a scalar mediator and $g_\chi^2/(4\pi)$ for the case
of a vector mediator.
For our study, it will be convenient to have one reference set of
parameters, and for this purpose we will use the values 
\beq
m_\chi = 5 \ {\rm GeV}, \quad m_\xi = 5 \ {\rm  MeV}, \ \quad
\alpha_\chi = 3 \times 10^{-4} \ , 
\label{parameter_values}
\eeq
where, as above, $\xi$ denotes $\phi$ or $V$ in the two respective
versions of the model. Thus, this model makes use of a light
mediator.  Motivations for this choice are discussed below.  We will
also calculate cross sections for a range of values of the coupling,
$\alpha_\chi$, and the mediator mass, $m_\xi$, and show how the results
compare with those obtained with the reference set of values in Eq.
(\ref{parameter_values}).  Note that the $\chi$ mass term is of Dirac form,
${\cal L}_{m_\chi} = m_\chi \bar\chi \chi$; we do not consider
Majorana mass terms for $\chi$ here.

Self-interacting dark matter models of this type have been shown to
ameliorate problems with excessive density on the scale of $\sim 1$
kpc in the cores of galaxies and to improve fits to morphological
properties of galaxies and, on larger length scales extending to
several Mpc, also improve fits observational data on clusters of
galaxies \cite{spergel_steinhardt},
\cite{dssw,kusenko_steinhardt,
mirror1,
bullet_cluster_constraints,
afsw,
feng_kaplinghat_yu,
buckley_fox,
koda,
vogelsberger_zavala_loeb,
kouvaris2012,
tulin_yu_zurek2013prl,
tulin_yu_zurek2013,
zurek_adm_review,
wise,
kusenko,
frandsen_sarkar,
elbert_bullock_peter,
kaplinghat_tulin_yu_prl2016,
slatyer_sommerfeld,
valli_reassessment,
kkpy,
battaglieri,
robertson2017,
robertson2018,
tulin_yu_review,
valli_yu,
lesgourgues_brinckmann_sidm,
brinckmann_vogelsberger,
vogelsberger_zavala_slatyer,
robertson2019,
rkky,
essig_sidm,
wechsler,
apr,
hayashi,
kaplinghat2020,
bondarenko,
essig_loverde,
tulin_velocity,
tulin_etal2021,
ebisu,
fischer2022a,
lisanti2022,
bullock_kaplinghat_valli2022,
eckert2022,
fischer2022}. 
Self-interacting dark matter models with scalar and/or vector
mediators are motivated by the fact that these yield DM-DM scattering
cross sections that decrease as a function of the relative velocities
$v_{\rm vel}$ of colliding DM particles, as is desirable to fit
observational data.  The reason for our
restriction to a vectorial gauge interaction in Eq. (\ref{chi_chi_v})
is that the generalization of this to a chiral gauge theory, with an
interaction ${\cal L} = q_L g [\bar\chi_L \gamma_\mu \chi_L]V^\mu +
q_R g [\bar\chi_R \gamma_\mu \chi_R]V^\mu$ in which the charges $q_L
\ne q_R$ would lead to triangle gauge anomalies unless one added
further DM fermions to cancel these.  To maintain maximal simplicity,
we have thus restricted this version of the model to the vectorial
interaction (\ref{chi_chi_v}).

The relative velocities of DM particles on all of the scales relevant
for galactic and cluster properties are nonrelativistic. Consequently,
an approach that has often been used is to model the scattering in
terms of a quantum-mechanical problem with a potential of the type
that would result in the nonrelativistic limit starting from the
$t$-channel exchange of the mediator.  In \cite{apr}, an analysis was
given of the full quantum field theoretic scattering of DM particles
in the case of reaction with incident $\chi + \bar\chi$. However,
Ref. \cite{apr} did not consider in depth the reaction
\beq
\chi + \chi \to \chi + \chi
\label{chichi_reaction}
\eeq
that is relevant to an ADM model.  In passing, we note that our
analysis is equally applicable for symmetric dark matter models;
however, in this case, the reaction (\ref{chichi_reaction}) only
contributes in part to the DM-DM scattering, the other process being
$\bar\chi + \chi \to \bar\chi + \chi$, which was considered extensively
in ref.~\cite{apr}. Here we focus on the reaction (\ref{chichi_reaction}). 


\section{Background}
\label{background_section}

In this section we explain the reasons for our choice of
parameter values (\ref{parameter_values}) in our model.  First, in
asymmetric dark matter models, with the asymmetries in the dark matter
and the baryons being of similar magnitude, it is plausible that 
\beq
\frac{m_{\rm DM}}{m_p} \simeq \frac{\rho_{_{\rm DM}}}{\rho_{b}} \simeq 5 \ , 
\label{mDM_adm}
\eeq
where $\rho_b$ is the average cosmological baryon density, and $m_p$
is the proton mass. This leads to the choice $m_\chi \simeq 5$ GeV.
(It should be noted that the simple relationship can be avoided in 
specific models, depending on the mechanisms that are assumed for
the generation of the $\chi$-$\bar\chi$ number asymmetry
\cite{zurek_adm_review}, but it will
suffice for our present purposes.) 
Second, as discussed above, SIDM fits to small-scale structure yield
$\sigma/m_{\rm DM} \sim 1$ cm$^2/$g.  Now, we will show that in our model, 
$\sigma/m_{\chi} \simeq {2 \pi \alpha_\chi^2 m_\chi}/{m_\xi^4}$. Setting this
equal to 1 cm$^2$/g determines the mediator mass $m_\xi$ to be 
\beq
m_\xi = \Big (\frac{\alpha_\chi}{1.2 \times 10^{-5}} \Big )^{1/2}
\Big( \frac{m_\chi}{5 \  {\rm GeV}}\Big)^{1/4} \ {\rm MeV} \ . 
\label{mxi_from_sig0}
\eeq
Third, in order to effectively annihilate away the symmetric
component of the dark matter in the early universe in the ADM model,
one requires a sizable
cross section for $\bar\chi \chi \to \xi \xi$. Note that, from
Eq. (\ref{mxi_from_sig0}), it follows that $m_\xi$ is naturally smaller than
$m_\chi$, so that this
process is kinematically allowed. The depletion of the symmetric component of the DM in the early Universe is satisfied when~\cite{zurek_adm_review, tulin_yu_zurek2013prl}
\beq
\langle \sigma v_{\rm rel} \rangle_{\bar\chi \chi \to \xi \xi} \simeq
\frac{\pi \alpha_\chi^2}{m_\chi^2} \sqrt{1- \frac{m_\xi^2}{m_\chi^2}}
\gtrsim 0.6 \times 10^{-25} \ {\rm cm^3/s} \ .
\label{adm_annihilation_constraint}
\eeq
Anticipating that $m_\xi \ll m_\chi$, this then yields a lower bound on
the SIDM coupling  strength, namely
\beq
\alpha_\chi \gtrsim 2 \times 10^{-4} \  \Big(\frac{m_\chi}{5 \ {\rm GeV}}\Big) \ .
\label{alpha_constraint}
\eeq
As stated before, for simplicity, we assume parameter values such that
lowest-order perturbative calculations are sufficiant to describe the
scattering.  From Eq. (\ref{born1}) in Appendix \ref{born_appendix},
this perturbativity condition requires that ${\alpha_\chi
  m_\chi}/{m_\phi} \ll 1$. Using the constraints in
Eqs. (\ref{mDM_adm}), (\ref{mxi_from_sig0}), (\ref{alpha_constraint}),
and (\ref{born1}), we then choose the values of the parameters in
eq. (\ref{parameter_values}).  Because the DM particle $\chi$ and the
mediator are SM-singlets, these choices for their masses are in accord
with bounds on DM particles and mediators from current data (for
summaries of bounds, see, e.g., \cite{mohapatra_nussinov_dmphoton,
  filippi_napoli,lanfranchi}).  Although we use the particular
set of values of the parameters in Eq. (\ref{parameter_values}) for much  
of our analysis, we also perform cross section calculations for a
substantial range of allowed values of $\alpha_\chi$ and $m_\chi$ in Section
\ref{parameter_variation_section}. These calculations show how our
results would change with different (allowed) values of parameters. 
Importantly, our choices for $m_\chi$ and $m_\xi$, which are motivated
from the above considerations, also lead to the desired velocity
dependences for the SIDM cross sections in the model that are of the
right order to fit observational data.

  
%
\section{Kinematics}
\label{kinematics_section}

In this section we review some basic kinematics relevant for our cross
section calculations.  Since the number density of $\bar\chi$ fermions
is much smaller than that of $\chi$ fermions after the $\bar\chi$
fermions have annihilated away in the ADM framework, the dominant
self-interactions of the $\chi$ DM particles arise from the reaction
(\ref{chichi_reaction}).  We take $\alpha_\chi$ to be sufficiently
small that the $\chi$-$\xi$ interaction can be well described by
lowest-order perturbation theory. This entails the condition that
there be no signicant Sommerfeld enhancement of the scattering.  In
the case of a vector mediator, the reaction (\ref{chichi_reaction})
involves a repulsive interaction of the $\chi$ particles, so there is
obviously no Sommerfeld enhancement.  Our choice of parameters
(\ref{parameter_values}) also guarantees the reliability of the
lowest-order perturbative calculation in the scalar case, as is
discussed further in Appendix \ref{born_appendix}.

\begin{figure*}
	\begin{center}
		\begin{subfigure}{0.42\textwidth}
			\centering
			\includegraphics[width=\textwidth]{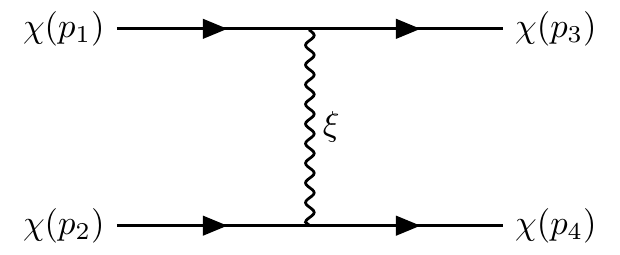}
			\subcaption{t-channel}
			\label{graph_tchannel}
		\end{subfigure}
		\hfill
		\begin{subfigure}{0.42\textwidth}
			\centering
			\includegraphics[width=\textwidth]{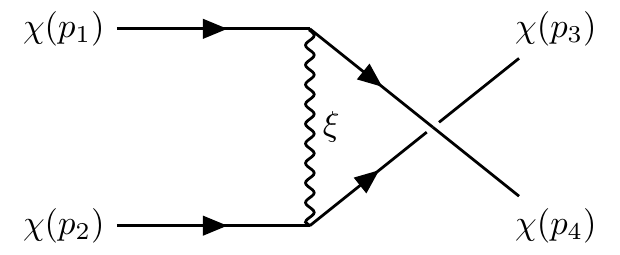}
			\subcaption{u-channel}
			\label{graph_uchannel}
		\end{subfigure}
	\end{center}
	\caption{Feynman diagrams for the reaction
          $\chi \chi \to \chi \chi$ via
          the exchange of the mediator particle, $\xi$.  We show the case where
          $\xi=V$. In standard notation, replacing the wavy line by a
          dashed line represents the case where $\xi=\phi$.}
	\label{graphs}
\end{figure*}

At tree level, there are two graphs contributing to the $\chi + \chi
\to \chi + \chi$ reaction, involving exchange of the mediator in the
$t$-channel and $u$-channel, with a relative minus sign between the two terms
in the amplitude, resulting from the fact that these two graphs are
related by the interchange of identical fermions in the final
state. These graphs and the associated momentum labelling are shown in
Fig. \ref{graphs}. For the reaction $\chi(p_1) + \chi(p_2) \to
\chi(p_3) + \chi(p_4)$, we define the usual invariants
\beqs
s &=& (p_1+p_2)^2 = (p_3+p_4)^2 \cr\cr
t &=& (p_1-p_3)^2 = (p_4-p_2)^2 \cr\cr
u &=& (p_1-p_4)^2 = (p_3-p_2)^2 \ . 
\label{stu}
\eeqs
We review some basic kinematics relevant for the analysis of this reaction.  
In the center-of-mass (CM) frame, the energies of each of the particles in
the initial and final states are the same and are equal to
\beq
E_\chi = \frac{\sqrt{s}}{2} \ .
\label{eif}
\eeq
Similarly, the magnitudes of the 3-momenta of each of the particles in
the initial and final states are the same
and are equal to
\beq
|{\vec p}_\chi| = \beta_\chi \frac{\sqrt{s}}{2} \ , 
\label{pif}
\eeq
where the magnitudes of the CM velocities are
\beq
\beta_\chi = \sqrt{1-\frac{4m_\chi^2}{s}} \ .
\label{beta}
\eeq
%
%
%
%
%

In the nonrelativistic limit, the relative velocity with which the two
$\chi$ particles approach each other is
\beq
\beta_{\rm rel} = 2\beta_\chi \ ,
\label{betarel}
\eeq
so in this limit, $|{\vec p}_\chi| = m_\chi \beta_{\rm rel}/2$. The
angle between ${\vec p}_1$ and ${\vec p}_3$ in the center of mass
frame is the CM scattering angle, $\theta$.  The invariants $s$, $t$,
and $u$ can be written in terms of $|{\vec p}_\chi|$ and $\theta$ as
\beqs
s &=& 4(m_\chi^2 + |{\vec p}_\chi|^2) \cr\cr
t &=& -4|{\vec p}_\chi|^2 \, \sin^2(\theta/2) \cr\cr
u &=& -4|{\vec p}_\chi|^2 \, \cos^2(\theta/2) \ .
\label{stu_values}
\eeqs
The transformation $\theta \to \pi - \theta$ interchanges the $t$ and
$u$ channels, as is evident in (\ref{stu_values}), since
$\sin[(1/2)(\pi-\theta)] = \cos(\theta/2)$.


\section{$\chi \chi \to \chi \chi$ Scattering Cross Sections with
  Scalar Mediator}
\label{scalar_mediator_section}


\subsection{Differential and Total Cross Sections}

The lowest-order (tree-level) amplitude for the $\chi + \chi \to \chi
+ \chi$ reaction resulting from the interaction (\ref{chi_chi_phi})
has the form
\beq
   {\cal M} = {\cal M}^{(t)} - {\cal M}^{(u)} \ , 
\label{amplitude}
\eeq
where ${\cal M}^{(t)}$ and ${\cal M}^{(u)}$ are the $t$-channel and
$u$-channel contributions, and the relative minus sign accounts for
exchanging identical fermions in the final state. The
Lorentz-invariant differential cross setion is
\beq
\frac{d\sigma}{dt} = 
 \frac{1}{16\pi \lambda(s,m_\chi^2,m_\chi^2)} \overline{\sum}
  |{\cal M}|^2 \ ,
  \label{dsigma_dt}
\eeq
where $\lambda(x,y,z) = x^2+y^2+z^2-2(xy+yz+zx)$, and $\overline{\sum}$ denotes
an average over initial spins and a sum over final spins.  Here, 
$\lambda(s,m_\chi^2,m_\chi^2) = (s\beta_\chi)^2$.  For our discussion,
it will be useful to distinguish the terms in $d\sigma/dt$ arising from
$\overline{\sum} |{\cal M}^{(t)}|^2$,
$\overline{\sum} |{\cal M}^{(u)}|^2$, and
$\overline{\sum}
[{\cal M}^{(t) *} {\cal M}^{(u)} +{\cal M}^{(u) *} {\cal M}^{(t)}] =
2\overline{\sum} {\rm Re}[{\cal M}^{(t) *} {\cal M}^{(u)}]$. We denote these as
$d\sigma^{(t)}/dt$,
$d\sigma^{(u)}/dt$, and 
$d\sigma^{(tu)}/dt$, respectively.
We find
\begin{widetext}
\begin{equation}
\frac{d\sigma}{dt} = \frac{\pi \alpha_\chi^2}{(\beta_\chi s)^2} \,
\bigg [ \frac{(t-4m_\chi^2)^2}{(t-m_\phi^2)^2} + 
  \frac{(u-4m_\chi^2)^2}{(u-m_\phi^2)^2} -
  \frac{1}{(t-m_\phi^2)(u-m_\phi^2)}\, \bigg \{ \frac{1}{2}(t^2+u^2-s^2)
  +8m_\chi^2 s - 8m_\chi^4 \bigg \} \bigg ] \ .
\label{dsigma_scalar}
\end{equation}
\end{widetext}
The first and second terms on the right-hand side (RHS)
of Eq. (\ref{dsigma_dt})
are $d\sigma^{(t)}/dt$ and $d\sigma^{(u)}/dt$, while the third term with curly
brackets is $d\sigma^{(tu)}/dt$. Since the amplitude (\ref{amplitude}) is
antisymmetric under interchange of identical particles in the final state,
and equivalently under interchange of the $t$-channel and $u$-channel terms, it
follows that the square of the amplitude is symmetric under this interchange.
This symmetry under the interchange $t \leftrightarrow u$ is evident
in the RHS of Eq. (\ref{dsigma_dt}). The center-of-mass
cross section, $(d\sigma/d\Omega)_{\rm CM}$, is related to $d\sigma/dt$ as
\beq
\bigg ( \frac{d\sigma}{d\Omega} \bigg )_{\rm CM} =
  \frac{\lambda(s,m_\chi^2,m_\chi^2)}{4\pi s} \,
\frac{d\sigma}{dt} = \bigg ( \frac{\beta_\chi^2 s}{4\pi} \bigg ) \,
\frac{d\sigma}{dt} \ .
\label{dsigma_domega_dsigma_dt}
\eeq
In terms of the center-of-mass scattering angle $\theta$, the symmetry of the
RHS of Eq. (\ref{dsigma_dt}) under the interchange $t \leftrightarrow u$ is
expressed as the symmetry
\beq
\bigg ( \frac{d\sigma}{d\Omega} \bigg )_{\rm CM}(\theta) =
\bigg ( \frac{d\sigma}{d\Omega} \bigg )_{\rm CM}(\pi-\theta) \ .
\label{dsigma_domega_sym}
\eeq
Because of the identical particles in the final state, a scattering
event in which a scattered $\chi$ particle emerges at angle $\theta$
is indistingishable from one in which a scattered $\chi$ emerges at
angle $\pi-\theta$. The total cross section for the reaction
(\ref{chichi_reaction})
thus involves a symmetry factor of $1/2$ to compensate for the
double-counting involved in the integration over the range
$\theta \in [0,\pi]$: 
\beq
\sigma = \frac{1}{2} \int d\Omega \,
\bigg ( \frac{d\sigma}{d\Omega} \bigg )_{\rm CM}(\pi-\theta) \ .
\label{sigma_eq}
\eeq
Owing to the symmetry (\ref{dsigma_domega_sym}), this is
equivalent to a polar angle integration from 0 to $\pi/2$:
\beq
\frac{1}{2}  \int_{-1}^1 \, d\cos\theta \, 
\bigg ( \frac{d\sigma}{d\Omega} \bigg )_{\rm CM} =
\int_0^1 \, d\cos\theta \, \bigg ( \frac{d\sigma}{d\Omega} \bigg )_{\rm CM} \ .
\label{hemisphere}
\eeq
(Recall that if the final state consisted of $n$ identical particles, the
factor $1/2$ in Eq. (\ref{sigma_eq}) would be replaced by $1/n!$.) 

In addition to the differential cross section
$(d\sigma/d\Omega)_{CM}$, other related (center-of-mass) differential
cross sections have been used in the study of the effects of
self-interacting dark matter, motivated by earlier analyses of
transport properties in gases and plasmas (e.g., \cite{ks} and
references therein).  A major reason for this was the desire to define
a differential cross section that yields a useful description of the
thermalization effect of DM-DM scattering, particularly in the case
where the mass of the mediator particle is much smaller than the mass
of the DM particle. In this case, to the extent that the scattering
angle $\theta$ is close to 0 for distinguishable particles or close to
0 or $\pi$ for indistinguishable particles, the DM particle
trajectories are not significantly changed by the scattering.  To give
greater weighting to large-angle scattering that thermalizes particles
in a gas or plasma, researchers \cite{ks} have used the transfer (T)
differential cross section,
\beq
\frac{d\sigma_{\rm T}}{d\Omega} = (1-\cos \theta)
\bigg ( \frac{d\sigma}{d\Omega} \bigg )_{\rm CM} 
\label{transfer_sigma}
\eeq
and the viscosity (V) differential cross section,
\beq
\frac{d\sigma_{\rm V}}{d\Omega} = (1-\cos^2 \theta)
\bigg ( \frac{d\sigma}{d\Omega} \bigg )_{\rm CM} \ ,
\label{viscosity_sigma}
\eeq
(Although the same symbol, V, is used for the vector mediator and viscosity,
the context will always make clear which meaning is intended.) 
For the same reason, namely that these describe thermalization effects
better than the ordinary cross section, the transfer and viscosity cross
sections been used in studies of DM-DM scattering (e.g.,
\cite{mirror1,tulin_yu_zurek2013} and subsequent work).

Given the invariance of $(d\sigma/d\Omega)_{CM}$ under the
transformation $\theta \to \pi - \theta$ and the fact that
$\cos\theta$ is odd under
this transformation, it follows that the integral of the product of
$\cos\theta$ times $(d\sigma/d\Omega)_{CM}$ vanishes. Hence, the total cross
section is equal to the total transfer cross section:
\beqs
\sigma &=& \frac{1}{2}
\int d\Omega \ \bigg ( \frac{d\sigma}{d\Omega} \bigg )_{\rm CM} 
= \frac{2 \pi}{2} \int_{-1}^1 \ d\cos\theta \,
\bigg ( \frac{d\sigma}{d\Omega} \bigg )_{\rm CM} \cr\cr
&=&  \frac{2 \pi}{2} \int_{-1}^1 \ d\cos\theta \, (1-\cos\theta)
\bigg ( \frac{d\sigma}{d\Omega} \bigg )_{\rm CM} = \sigma_{\rm T} \ .
\cr\cr
&&
\label{sig_sigtransfer_equality}
\eeqs
As was noted in \cite{ks}, the transfer differential cross section
does not correctly describe the scattering in the case of identical
particles, since it does not maintain the $\theta \leftrightarrow \pi
- \theta$ symmetry in the reaction.  However, given
Eq. (\ref{sig_sigtransfer_equality}), the resultant integral over
angles is equal to the integral of the ordinary (unweighted) cross
section, i.e., $\sigma_T = \sigma$.  The viscosity differential cross
section, with its angle-weighting factor of
$(1-\cos^2\theta) = \sin^2\theta$
does maintain the $\theta \leftrightarrow \pi - \theta$
symmetry in the scattering of identical particles. In passing, we note
that another type of differential cross section has also been considered
that weights large-angle scattering
\cite{frandsen_sarkar}, namely
$(1-|\cos\theta|)(d\sigma/d\Omega)_{\rm CM}$; this also maintains the
$\theta \to \pi - \theta$ symmetry of reaction (\ref{chichi_reaction}).) 

In the nonrelativistic (NR) limit $\beta_\chi \ll 1$, the kinematic
invariants have the property that $s \gg \{ |t|, \ |u|\}$;  $m_\chi^2 \gg
\{ |t|, \ |u|\}$; and $s \to (2m_\chi)^2$. Hence, in this limit, the CM
differential cross section reduces to
\begin{widetext}
\beqs
\bigg ( \frac{d\sigma}{d\Omega} \bigg )_{\rm CM,NR} &=&
  {\alpha_\chi^2 m_\chi^2} \bigg [
\frac{1}{(t-m_\phi^2)^2} + \frac{1}{(u-m_\phi^2)^2} -
\frac{1}{(t-m_\phi^2)(u-m_\phi^2)} \bigg ] \cr\cr
&=& {\sigma_0} \bigg [
\frac{1}{(1+r\sin^2(\theta/2))^2} +
\frac{1}{(1+r\cos^2(\theta/2))^2} -
\frac{1}{(1+r\sin^2(\theta/2))(1+r\cos^2(\theta/2))} \bigg ] \ ,  \cr\cr 
 &&
\label{dsigma_domega_nonrel}
\eeqs
\end{widetext}
where 
\beq
\sigma_0 = \frac{\alpha_\chi^2  m_\chi^2}{m_\phi^4} 
\label{sigma0}
\eeq
and $r$ is the ratio
\beq
r = \bigg ( \frac{\beta_{\rm rel} m_\chi}{m_\phi} \bigg )^2 \ .
\label{r}
\eeq
The property that the transformation $\theta \to \pi - \theta$ (under
which $\sin(\theta/2) \to \cos(\theta/2)$) interchanges the $t$ and
$u$ channels is evident in Eq. (\ref{dsigma_domega_nonrel}), since it
interchanges the first and second terms arising, respectively, from
$|{\cal M}^{(t)}|^2$ and from $|{\cal M}^{(u)}|^2$, and leaves the
third term arising from $-2{\rm Re}({\cal M}^{(t)*}{\cal M}^{(u)})$
invariant.  Since all of the $\chi$-$\chi$ relative velocities $v_{\rm rel}$
in the relevant observational data are nonrelativistic, we will henceforth
specialize to this case, taking the subscript NR to be implicit in the
notation. 

Since self-interacting dark matter has been studied extensively
before, it is appropriate to discuss how our current results
compare with and complement previous work.  In (Eq. (25) of) the review
\cite{tulin_yu_review} on SIDM, the differential cross section in the
center of mass for elastic DM self-scattering was given (in the same
perturbative Born regime $\alpha_\chi m_\chi/m_\phi \ll 1$ as we use
here) as
\beq
\frac{d\sigma}{d\Omega} = \frac{\alpha_\chi^2 m_\chi^2}
     {[m_\chi^2 v_{\rm rel}^2(1-\cos\theta)/2 + m_\phi^2]^2} \equiv
\frac{\sigma_0}{ [r\sin^2(\theta/2) + 1]^2}  \ , 
\label{dsigma_domega_tulinyu}
\eeq
where we transcribe the result from \cite{tulin_yu_review} in our
notation in the second term of Eq. (\ref{dsigma_domega_tulinyu}).  As
is evident, this corresponds to the $t$-channel contribution in our
full result (\ref{dsigma_domega_nonrel}). However, the true
differential cross section for the DM self-scattering $\chi + \chi \to
\chi + \chi$ must include not just the $t$-channel contribution but
also the $u$-channel contribution, as we have done here.  A subsequent
study in \cite{tulin_etal2021} focused on a regime where
nonperturbative effects are important and gave results for DM-DM
scattering with both identical and non-identical particles.  Our work
is complementary to \cite{tulin_etal2021}, since we choose parameters
in Eq. (\ref{parameter_values}) such that nonperturbative effects are
not important.

Regarding the range of values of the ratio $r$ in Eq. (\ref{r}), it is
important to note that even in the nonrelativistic regime $\beta_{\rm
  rel} \ll 1$, it is not necessarily the case that the ratio $r$ is
small. With the illustrative mass values in
Eq. (\ref{parameter_values}), and taking into account that for $v_{\rm
  rel} \sim 3 \times 10^3$ km/s (i.e., $\beta_{\rm rel} \sim 10^{-2}$)
for DM particles in galaxy clusters, it follows that $r \sim 10^2$ in
this case.  In contrast, for the analysis of DM self-interactions on
length scales of order a few kpc within a galaxy, if $v_{\rm rel} \sim
30$ km/sec (i.e., $\beta_{\rm rel} \sim 10^{-4}$), then $r \sim
O(10^{-2})$.

It is interesting to elucidate how the various contributions to the
cross section from $|{\cal M}^{(t)}|^2$, $|{\cal M}^{(u)}|^2$, and
$2{\rm Re}({\cal M}^{(t)*}{\cal M}^{(u)})$ behave as a function of
$r$. We find that in the $r \ll 1$ regime relevant for the analysis
of galactic data on the 1-10 kpc scale, the terms contributing to
$(d\sigma/d\Omega)_{\rm CM}$ have the property that the $t$-channel
term $|{\cal M}^{(t)}|^2$ and the $u$-channel term $|{\cal
  M}^{(u)}|^2$ give equal contributions, while the $t$-$u$
interference term $2{\rm Re}({\cal M}^{(t)*}{\cal M}^{(u)})$ gives a
contribution equal in magnitude and opposite in sign to that from
$|{\cal M}^{(u)}|^2$. As we denoted the three terms contributing to
$d\sigma/dt$, we similarly label the three terms contributing to
$(d\sigma/d\Omega)_{\rm CM}$ and the resultant total cross section
with superscripts $(t)$, $(u)$, and $(tu)$, so that the respective
contribution to the total cross section are
\beq
\sigma^{(i)} \equiv \frac{1}{2} 
\int \bigg ( \frac{d\sigma^{(i)} }{d\Omega} \bigg )_{\rm CM} \,
d\Omega \ , \quad i=t, \ u, \ tu \ ,
\label{sigmai_nonrel}
\eeq
and
\beq
\sigma = \sigma^{(t)}+\sigma^{(u)}+\sigma^{(tu)} \ . 
\label{sigma_sum_nonrel}
\eeq
We calculate
\beq
\sigma^{(t)} = \sigma^{(u)} = \frac{2\pi\sigma_0}{1+r} 
\label{sigmat_sigmau_nonrel}
\eeq
and
\beq
\sigma^{(tu)} = -4\pi\sigma_0 \, \frac{\ln(1+r)}{r(2+r)} \ ,  
\label{sigmatu_nonrel}
\eeq
so that
\beq
\sigma = 4\pi \sigma_0 \bigg [ \frac{1}{1+r} - \frac{\ln(1+r)}{r(2+r)}
  \bigg ] \ .
\label{sigma_nonrel}
\eeq
For fixed $\sigma_0$, the total cross section $\sigma$ is a
monotonically decreasing function of the ratio $r$. 
Concerning the individual contributions to $\sigma$, we observe that
\beq
\sigma^{(t)} = \sigma^{(u)} = - \sigma^{(tu)} =
2\pi\sigma_0 \quad {\rm at} \ r=0 \ , 
\label{sigt_sigu_r_0}
\eeq
so that for small $r$, there is a cancellation between the interference
term $\sigma^{(tu)}$ and the $u$-channel term $\sigma^{(u)}$ (or equivalently,
the $t$-channel term, since $\sigma^{(t)}=\sigma^{(u)}$). 
In contrast, for large $r$, $\sigma^{(t)} = \sigma^{(u)}$ decrease as
$2\pi\sigma_0/r$, while $\sigma^{(tu)}$ decreases more rapidly, as
$\sigma^{(tu)} \sim -4\pi\sigma_0 \ln r/r^2$.  The total cross section
has the small-$r$ Taylor series expansion 
\beq
\sigma  = 2\pi \sigma_0 \Big [ 1 -r + \frac{7}{6}r^2 +O(r^3) \Big ]
\quad {\rm for} \ r \ll 1 \ .
\label{sigma_nonrel_small_r}
\eeq
For $r \gg 1$, $\sigma$ has the series expansion 
\beqs
\sigma &=& \frac{4\pi \sigma_0}{r} \bigg [ 1 -\frac{(1+\ln r)}{r}
  + O\bigg (\frac{\ln r}{r^2}\bigg ) \bigg ] \cr\cr
&& {\rm for \ NR \ regime \ and } \ r \gg 1 \ . 
\label{sigma_nonrel_large_r}
\eeqs
The prefactor in Eq. (\ref{sigma_nonrel_large_r}) is
\beq
\frac{4\pi\sigma_0}{r} =
\frac{4\pi \alpha_\chi^2}{m_\phi^2 \beta_{\rm rel}^2} \ .
\label{prefactor1}
\eeq
To compare the full cross section with the result obtained by
including only the contribution from the $t$-channel, we consider the ratio
\beqs
\frac{\sigma}{\sigma^{(t)}} =
2\bigg [ 1 - \frac{(1+r)\ln(1+r)}{r(2+r)} \bigg ] \ . 
\label{sig_over_sigt}
\eeqs
This ratio has the small-$r$ expansion 
\beq
\frac{\sigma}{\sigma^{(t)}} =
1 + \frac{r^2}{6} - \frac{r^3}{6} + O(r^4) \quad {\rm for} \ r \ll 1 \ , 
\label{sigratio_small_r}
\eeq
so in the small-$r$ regime, $\sigma$ is approximately equal to
$\sigma^{(t)}$. For the (nonrelativistic) large-$r$ regime, the ratio
(\ref{sig_over_sigt}) has the expansion
\beqs
\frac{\sigma}{\sigma^{(t)} } &=&
2\bigg [ 1 - \frac{\ln r}{r} + \frac{\ln r - 1}{r^2} +
  O\bigg ( \frac{\ln r}{r^3} \bigg ) \bigg ] \cr\cr
&& {\rm for \ NR \ regime \ and} \ r \gg 1 \ .
  \label{sigratio_large_r}
\eeqs
Thus, in this large-$r$ regime relevant for fits to observational data
on galaxy clusters, the full $\chi$-$\chi$ scattering cross section is
larger by approximately a factor of 2 then the result obtained by keeping
only the contribution from the $t$-channel.

\begin{figure*}[htb!]
	\begin{center}
		\begin{subfigure}{0.46\textwidth}
			\centering
			\includegraphics[width=\textwidth]{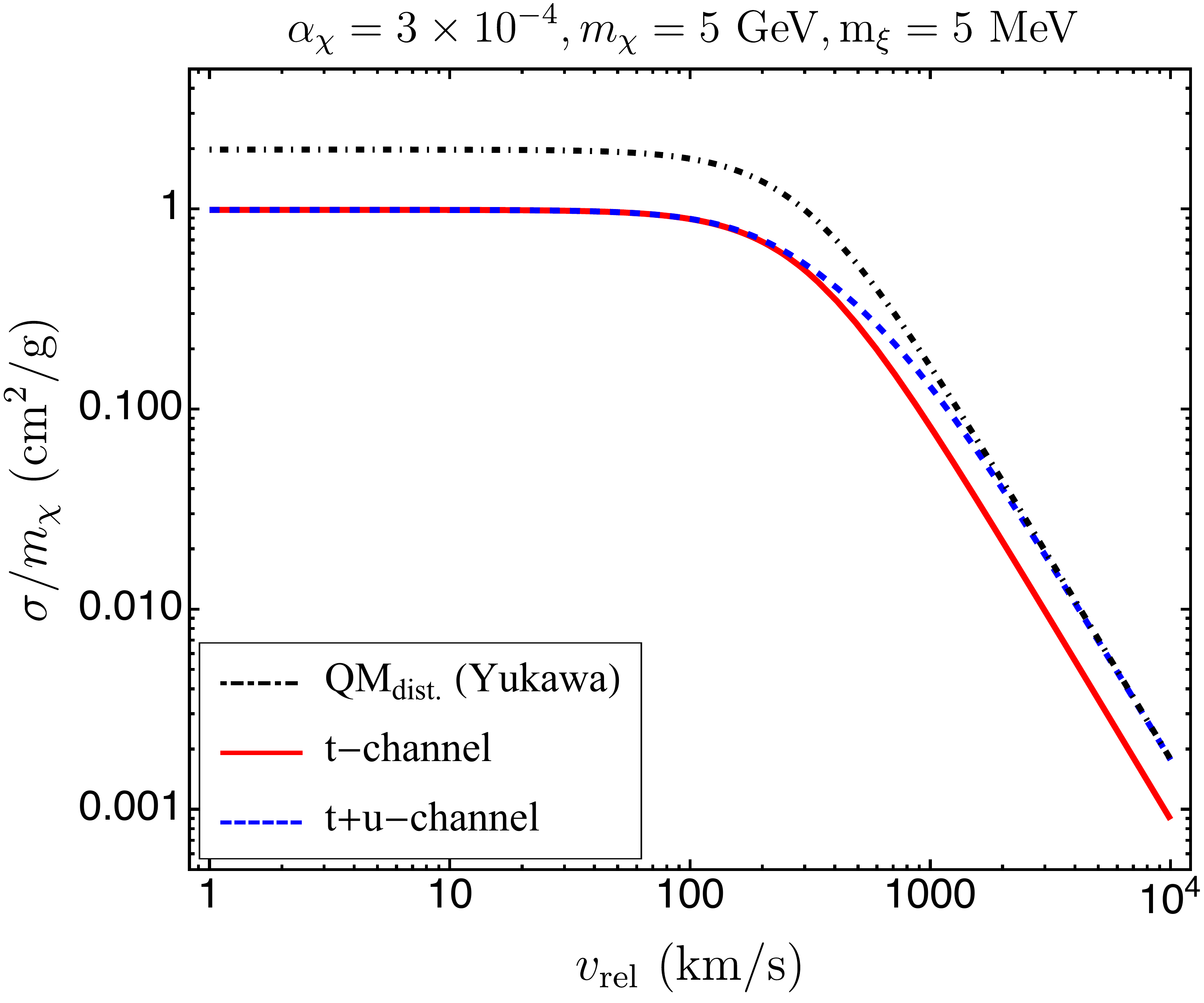}
			\subcaption{Cross section}
			\label{sidm_COM_comp_figure}
		\end{subfigure}
		\hfill
		\begin{subfigure}{0.46\textwidth}
			\centering
			\includegraphics[width=\textwidth]{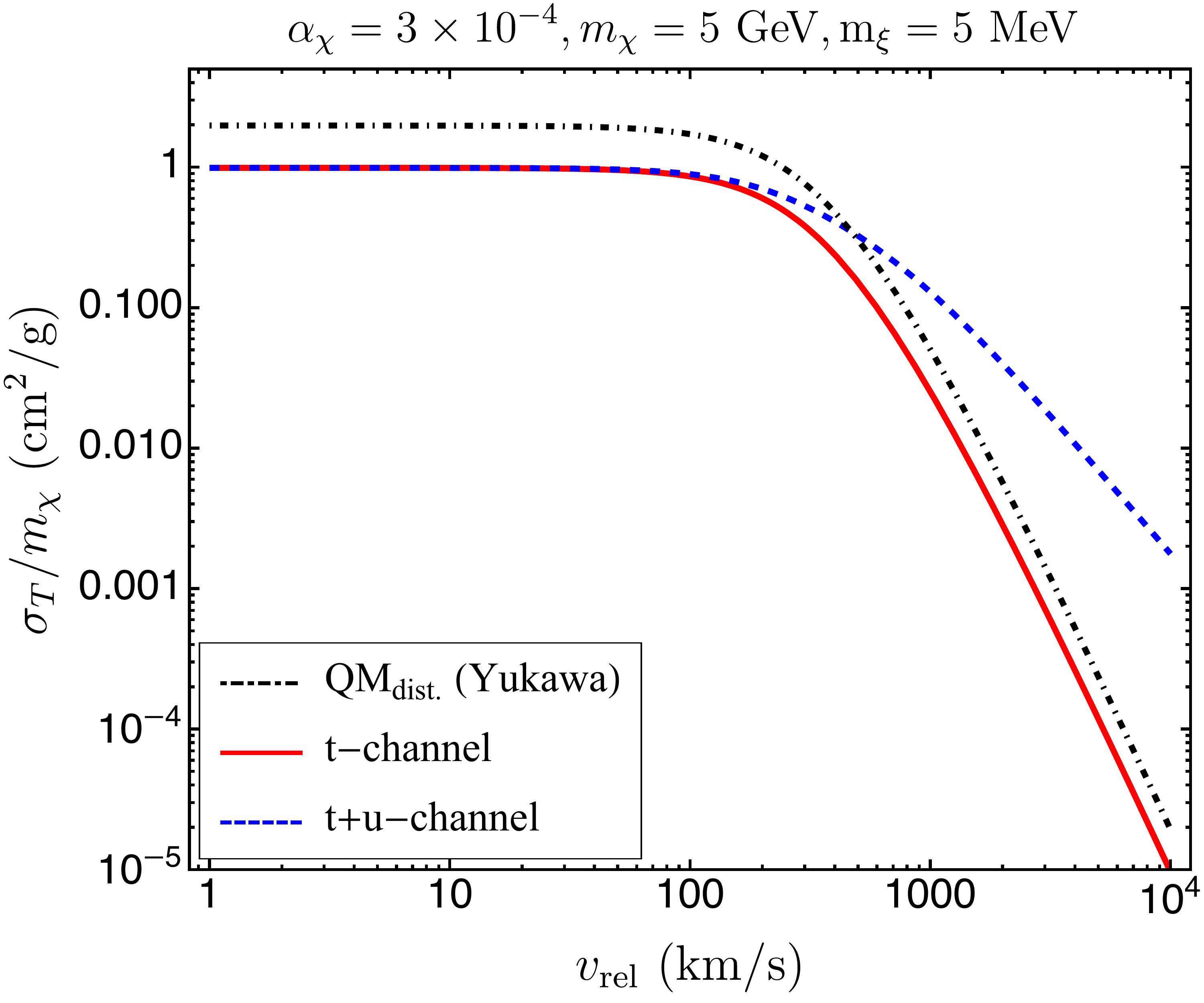}
			\subcaption{Transfer cross section}
			\label{sidm_transfer_comp_figure}
		\end{subfigure}
		\hfill
		\begin{subfigure}{0.46\textwidth}
			\centering
			\includegraphics[width=\textwidth]{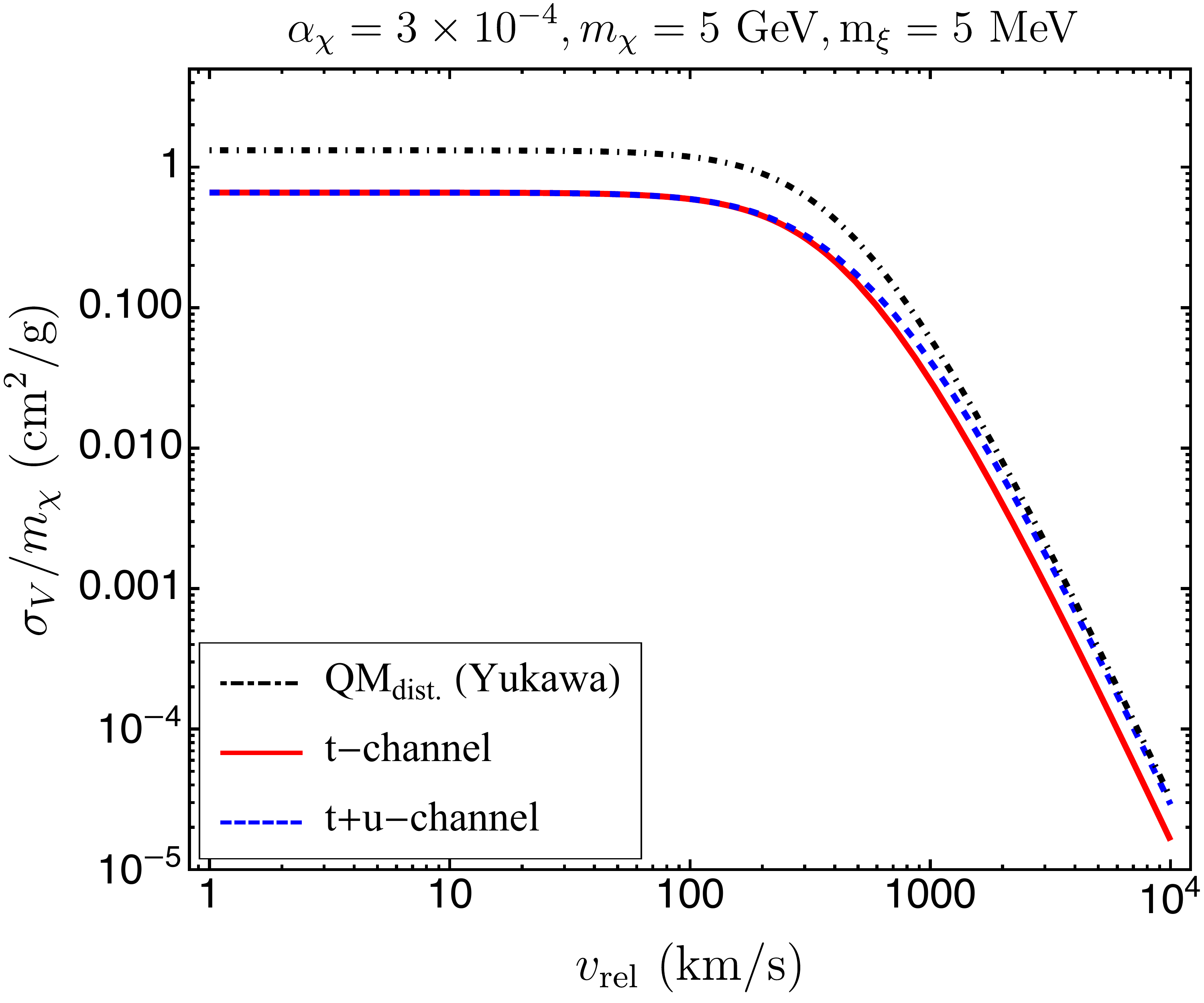}
			\subcaption{Viscosity cross section}
			\label{sidm_viscosity_comp_figure}
		\end{subfigure}
		\hfill
		\begin{subfigure}{0.46\textwidth}
			\centering
			\includegraphics[width=\textwidth]{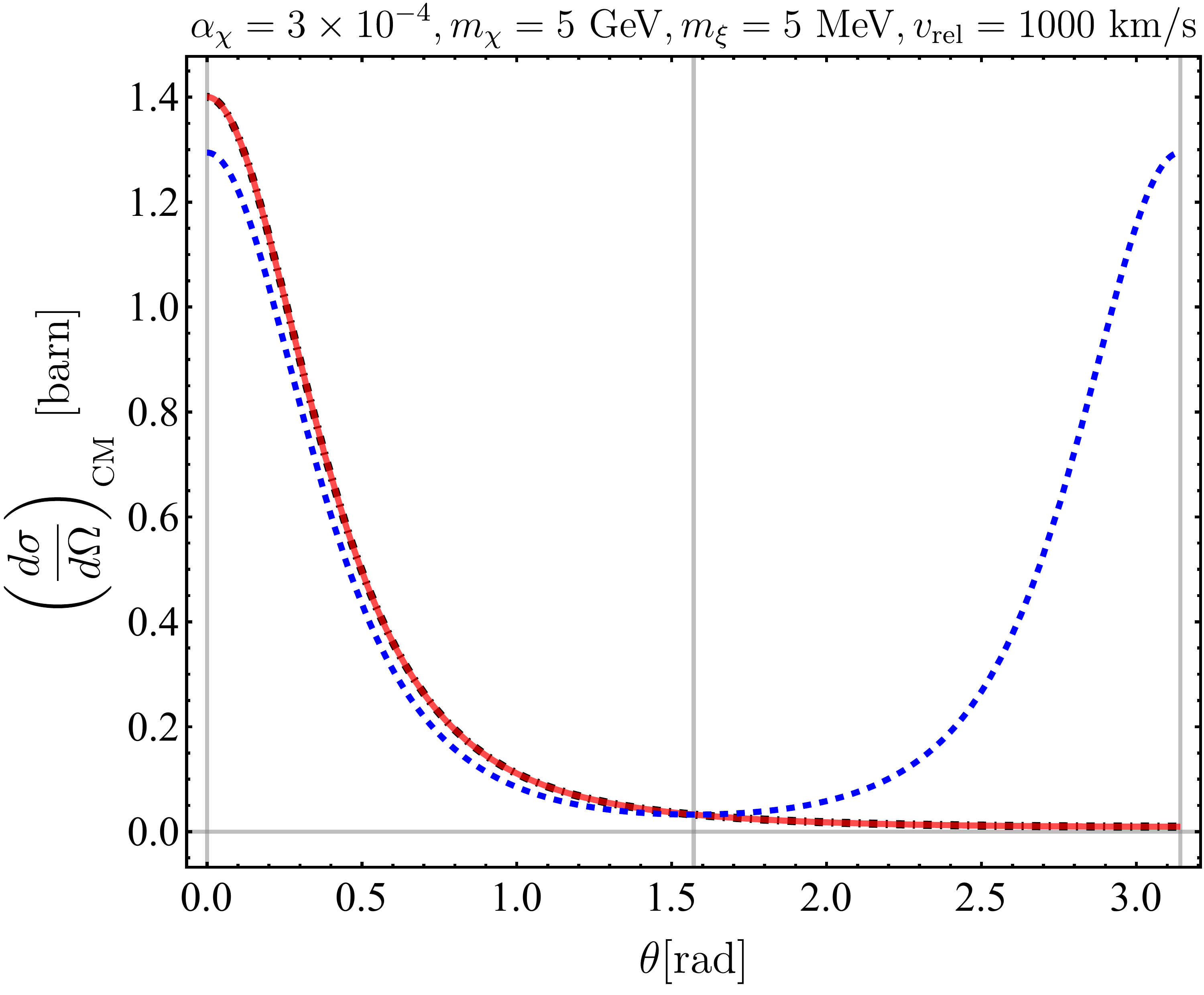}
			\subcaption{Dependence of differential scattering
                          cross section on the scattering angle}
			\label{sidm_angle_comp}
		\end{subfigure}
	\end{center}
	\caption{\footnotesize Fig. \ref{sidm_COM_comp_figure}
          shows $\sigma/m_\chi$ for the reaction $\chi + \chi \to \chi
          + \chi$, as a function of the relative velocity $v_{\rm
            rel}$ of the colliding $\chi$ particles.  The full result
          with proper inclusion of both $t$-channel and $u$-channel
          contributions is shown as the dashed curve (colored blue
          online), while the result of including only the $t$-channel,
          is indicated by the solid curve (colored red online).  
          The curves marked QM$_{\rm dist}$ refer to the
          result that one would get in a quantum mechanical approach
          to the different situation with distinguishable particles
          (see Appendix). The illustrative values $m_\chi=5$ GeV,
          $m_\xi=5$ MeV, and $\alpha_\chi=3 \times 10^{-4}$ given in
          Eq.  (\ref{parameter_values}) are used for the calculation.
          Fig. \ref{sidm_transfer_comp_figure} and Fig. 
          \ref{sidm_viscosity_comp_figure}  present the corresponding
          plots of the transfer and viscosity cross sections
          respectively. Fig. \ref{sidm_angle_comp} shows the
          dependence of the differential CM scattering
          cross section on the scattering angle. The color coding in
          Fig. \ref{sidm_angle_comp} is the same as in the other figures.}
	\label{sidm_comp_figs}
\end{figure*}

In order to compare the full calculation including contributions from
both the $t$-channel and $u$-channel with a calculation that only
includes the $t$-channel, we plot $\sigma$ versus $\sigma^{(t)}$ in
Fig. \ref{sidm_comp_figs} as a function of $v_{\rm rel}$. For this
purpose, we use the illustrative values of masses and couplings in
Eq. (\ref{parameter_values}).  In accordance with our result
(\ref{sigma_scalar_vector_relation}) below, we subsume the cases of a
scalar and a vector mediator together and denote $m_{\xi}$ as the mass
of $\phi$ or $V$.  We note again that with these values, there is no
significant Sommerfeld enhancement of the cross section, justifying
our use of the lowest-order (tree-level) perturbatively computed
amplitude in the scalar case. Separately, there is no Sommerfeld
enhancement in the vector case since the scattering is repulsive.  The
dependence of the differential cross sections on the angle $\theta$ is
shown in the comparative Fig. \ref{sidm_angle_comp}.  As is evident
from Fig. \ref{sidm_comp_figs}, for the range of relative velocities
$v_{\rm rel} \lsim 10^2$ km/s relevant for dark matter scattering in
the interior of galaxies and dwarf spheroidal satellites, $\sigma$ is
close to $\sigma^{(t)}$, but as $v_{\rm rel}$ increases beyond about
$10^2$ km/s, although both $\sigma$ and $\sigma^{(t)}$ decrease, the
full cross section is larger than the result obtained by keeping only
the $t$-channel contribution.  This trend continues to values $v_{\rm
  rel} \sim O(10^3)$ km/s relevant to dark matter effects in galaxy
clusters.  One should note that even for a fixed $v_{\rm rel}$, there
is considerable diversity in the values of $\sigma/m_{DM}$ inferred
from fits to galactic and cluster data (e.g.,
\cite{kkpy,robertson2018,lisanti2022,roper_frenk_navarro} and
references therein).  The curves marked QM$_{\rm dist}$ in
Fig. \ref{sidm_comp_figs} are the results that one would obtain in a
quantum mechanical approach with a potential for the different
situation with distinguishable particles (see Appendix).  We show
results for a specific set of $v_{\rm rel}$ values in Table
\ref{table_COM_comp}.

%
%



\begin{table*}
\begin{center}
\begin{tabular}{|c|c|c|c|c|c|c|} 	 \hline\hline
$v_{\rm rel}$ \ (km/s) &
$\sigma^{(t)}/m_\chi$ \ (cm$^2$/g) &
$\sigma/m_\chi$         &
$\sigma^{(t)}_{\rm T}/m_\chi$ &
$\sigma_{\rm T}/m_\chi$ &
$\sigma^{(t)}_{\rm V}/m_\chi$ &
$\sigma_{\rm V}/m_\chi$ \\
\hline
$10$   & 0.99 & 0.99 & 0.99 & 0.99 & 0.66 & 0.66 \\
\hline
$10^2$ & 0.90 & 0.90 & 0.86 & 0.89 & 0.59 & 0.59 \\
\hline
$10^3$ & 0.082& 0.13 &0.025 & 0.13 & 0.030 & 0.042 \\
\hline
$10^4$ & $0.89 \times 10^{-3}$ & $1.8  \times 10^{-3}$
       & $0.96 \times 10^{-5}$ & $1.8 \times 10^{-3}$ 
       & $1.6 \times 10^{-5}$  & $2.9 \times 10^{-5}$  \\
\hline\hline
\end{tabular}
\end{center}
\caption{\footnotesize Comparison of different cross sections divided
  by dark matter particle mass, $m_\chi$, in units of cm$^2$/g,
  as functions of
          $v_{\rm rel}$. The calculations use the parameter values in
          Eq. (\ref{parameter_values}). See text for further details. }
	\label{table_COM_comp}
\end{table*}


\subsection{Transfer Cross Sections}
\label{transfer_section}

Our result in Eq. (\ref{dsigma_domega_nonrel}) together with the definition
(\ref{transfer_sigma}) yields the differential transfer cross section
in the relevant nonrelativistic limit. For the individual contributions
from $|{\cal M}^{(t)}|^2$, $|{\cal M}^{(u)}|^2$, and 
$-2{\rm Re}({\cal M}^{(t)*}{\cal M}^{(u)})$, we calculate (in the
nonrelativistic regime, as before), 
\beq
\sigma^{(t)}_{\rm T} = \frac{4\pi\sigma_0}{r}
\bigg [ -\frac{1}{1+r} + \frac{\ln(1+r)}{r} \bigg ] \ , 
\label{sigma_t_transfer_nonrel}
\eeq
\beq
\sigma^{(u)}_{\rm T} = \frac{4\pi\sigma_0}{r}
\bigg [ 1 - \frac{\ln(1+r)}{r} \bigg ] \ , 
\label{sigma_u_transfer_nonrel}
\eeq
and
\beq
\sigma^{(tu)}_{\rm T} = -\frac{4\pi\sigma_0}{r}
\bigg [ \frac{\ln(1+r)}{2+r} \bigg ] \ .
\label{sigma_tu_transfer_nonrel}
\eeq
The prefactor in
Eqs. (\ref{sigma_t_transfer_nonrel})-(\ref{sigma_tu_transfer_nonrel})
is given by Eq. (\ref{prefactor1}).  Note that, in contrast to the
equality $\sigma^{(t)}=\sigma^{(u)}$ in Eq. (\ref{sigmat_sigmau_nonrel}), the
individual contributions $\sigma^{(t)}_{\rm T}$ and $\sigma^{(u)}_{\rm T}$
to $\sigma_{\rm T}$
are not equal; i.e., $\sigma^{(t)}_{\rm T} \ne
\sigma^{(u)}_{\rm T}$.  This is a consequence of the fact that the
definition of $d\sigma_{\rm T}/d\Omega$ fails to preserve the
$\theta \to \pi - \theta$ symmetry of the actual differential cross
section for the reaction (\ref{chichi_reaction}). .

Summing these contributions, we find, in accordance with our general equality
(\ref{sig_sigtransfer_equality}), the result 
\beqs
\sigma_{\rm T} &=& \sigma =
4\pi \sigma_0 \bigg [ \frac{1}{1+r} - \frac{\ln(1+r)}{r(2+r)} \bigg ] \ .
\cr\cr
&&
\label{sigma_sigma_transfer_equality}
\eeqs
Since $\sigma_{\rm T} = \sigma$, the transfer cross section has
the same small-$r$ and large-$r$ expansions as were displayed for
$\sigma$ in Eqs. (\ref{sigma_nonrel_small_r}) and
(\ref{sigma_nonrel_large_r}).

We may compare our result (\ref{sigma_sigma_transfer_equality}) for
$\sigma_{T}$ with the result given, in the same Born regime, as
Eq. (A1) in Ref. \cite{tulin_yu_zurek2013prl} (denoted TYZ), which is the same
as Eq. (5) in Ref. \cite{feng_kaplinghat_yu} (denoted FKY) and reads (with
their $R \equiv \sqrt{r}$ and $v=\beta_{\rm rel}$ in our notation)
\beqs
\sigma_{T;FKY,TYZ}
&=& \frac{8\pi\alpha_\chi^2}{m_\chi^2 \beta_{\rm rel}^4}
\bigg [\ln(1+r) - \frac{r}{(1+r)} \biggl ]  \cr\cr
&=& \frac{8\pi\alpha_\chi^2 r}{m_\chi^2 \beta_{\rm rel}^4}
\bigg [\frac{\ln(1+r)}{r} -\frac{1}{1+r} \biggl ]  \cr\cr
&=& \frac{8\pi\alpha_\chi^2}{m_\phi^2\beta_{\rm rel}^2}
\bigg [\frac{\ln(1+r)}{r} -\frac{1}{1+r} \biggl ]  \ .
\cr\cr
&&
      \label{sigma_transfer_tyz}
\eeqs
As is evident from a comparison of Eq. (\ref{sigma_transfer_tyz}) with
our Eq. (\ref{sigma_t_transfer_nonrel}) (using the definition of our
notation given in Eq. (\ref{prefactor1})), the result for the transfer
cross section in Eq. (A1) of Ref. \cite{tulin_yu_zurek2013prl} (or
equivalently, Eq. (5) of Ref. \cite{feng_kaplinghat_yu}) is what one
would get for the DM self-scattering if one did the calculation
for non-identical particles and
hence only included the $t$-channel contribution and did not include
the $1/2$ factor for identical particles in the final state in
performing the integral over $d\Omega$. That is,
\beq
\sigma_{\rm T,TYZ,FKY} = 2\sigma^{(t)}_{\rm T} \ .
\label{sigt_tyz_versus_our_sigt}
\eeq

To compare the full transfer cross section with the result obtained 
by just including the $t$-channel contribution, we examine the ratio
\beq
\frac{\sigma_{\rm T}}{\sigma^{(t)}_{\rm T}} =
\frac{r \Big [ \frac{1}{1+r} - \frac{\ln(1+r)}{r(2+r)} \Big ]}
     {\Big [ - \frac{1}{1+r} + \frac{\ln(1+r)}{r} \Big ]} \ .
\label{sigma_transfer_over_sigmat_transfer}
\eeq
For small $r$, this ratio has the expansion
\beqs
\frac{\sigma_{\rm T}}{\sigma^{(t)}_{\rm T}} = 
  1 + \frac{r}{3} +
  \frac{r^2}{9} + O(r^3) \quad {\rm for} \ r \ll 1 \ .
\cr\cr
&&
\label{sigma_transfer_over_sigmat_transfer_small_r}
\eeqs
For large $r$, we find 
\beq
\frac{\sigma_{\rm T}}{\sigma^{(t)}_{\rm T}} \sim \frac{r}{\ln r-1} 
\quad {\rm for} \ r \gg 1 \ . 
\label{sigma_transfer_over_sigmat_transfer_large_r}
\eeq
Thus, although both our $\sigma_{\rm T}$ and the result $\sigma_{\rm
  T,FKY,TYZ}$ decrease with $v_{\rm rel}$ (and thus with $r$, for
fixed $m_\chi$ and $m_\xi$), our result decreases substantially less
rapidly for large $r$. With our parameters, this large-$r$ regime
includes values of $v_{\rm rel} \sim O(10^3)$ km/s typical of galaxy
clusters.  For example, at $v_{\rm rel} = 3 \times 10^3$ km/s
(corresponding to $r=10^2$ with our choices for $m_\chi$ and $m_\xi$
in Eq. (\ref{parameter_values})), the ratio
(\ref{sigma_transfer_over_sigmat_transfer}) has the value 26, or
equivalently, $\sigma_{\rm T}/\sigma_{\rm T,FKY,TYZ} = 13$, a
substantial difference from unity.  Therefore, in performing fits to
observational data, if one uses the transfer cross section, we would
advocate the use of Eq. (\ref{sigma_sigma_transfer_equality}) rather
than the result in Eq. (A1) of Ref. \cite{tulin_yu_zurek2013prl} for
the large-$r$ regime, since they differ substantially.

In Fig. \ref{sidm_transfer_comp_figure} we plot $\sigma_{\rm T}$ in
comparison with $\sigma^{(t)}_{\rm T}$ over the same range of $v_{\rm
  vel}$ and thus also $\beta_{\rm rel}$ as for the regular CM cross
section. The fact that the true $\sigma_{\rm T}$ decreases
considerably less rapidly than the $t$-channel contribution used in
\cite{feng_kaplinghat_yu,tulin_yu_zurek2013prl} is evident in this
figure.  This is also apparent in Table \ref{table_COM_comp}.  


\subsection{Viscosity Cross Section}

For the viscosity cross section we calculate the following contributions
from the $t$-channel, $u$-channel, and $t$-$u$ interference:
\beqs
\sigma^{(t)}_{\rm V} &=& \sigma^{(u)}_{\rm V} \cr\cr
&=& \frac{8\pi\sigma_0}{r^2}\bigg [ -2 + (2+r)\frac{\ln(1+r)}{r} \bigg ]
\label{sigma_visc_t_nonrel}
\eeqs
and
\beq
\sigma^{(tu)}_{\rm V} = \frac{8\pi\sigma_0}{r^2}\bigg [
  -1 + \frac{2(1+r)\ln(1+r)}{(2+r)r} \bigg ] \ ,
\label{sigma_visc_tu_nonrel}
\eeq
so that the total nonrelativistic viscosity cross section is
\beqs
\sigma_{\rm V} &=&
\sigma^{(t)}_{\rm V}+
\sigma^{(u)}_{\rm V}+
\sigma^{(tu)}_{\rm V} \cr\cr
&=& \frac{8\pi\sigma_0}{r^2}\bigg [
  -5 + \frac{2(5+5r+r^2)\ln(1+r)}{(2+r)r} \bigg ]
\ .
\cr\cr
&&
\label{sigma_visc_nonrel}
\eeqs
As was the case with $\sigma$ and $\sigma_{\rm T}$,
for fixed $m_\chi$ and $m_\phi$, the viscosity cross section
$\sigma_{\rm V}$ is a monotonically decreasing function of $r$.

We remark on properties of the individual contributions
$\sigma^{(t)}_{\rm V}$, $\sigma^{(u)}_{\rm V}$, and $\sigma^{(tu)}_{\rm V}$.
The fact that $\sigma^{(t)}_{\rm V}=\sigma^{(u)}_{\rm V}$ is guaranteed by
the property that $(d\sigma/d\Omega)_{\rm V}$ maintains the
$\theta \to \pi - \theta$ symmetry of $(d\sigma/d\Omega)_{\rm CM}$, which
interchanges the $t$- and $u$-channels.  These contributions
have the small-$r$ expansions
\beqs
\sigma^{(t)}_{\rm V} &=& \sigma^{(u)}_{\rm V} = \frac{4\pi\sigma_0}{3}
\bigg [ 1 - r + \frac{9}{10}r^2 + O(r^3) \bigg ] \cr\cr
&& {\rm for \ } r \ll 1 \
\label{sigt_visc_small_r}
\eeqs
and
\beqs
\sigma^{(tu)}_{\rm V} &=& \sigma^{(u)}_{\rm V} = \frac{4\pi\sigma_0}{3}
\bigg [ -1 + r - \frac{4}{5}r^2 + O(r^3) \bigg ] \cr\cr
&& {\rm for \ } r \ll 1 \ .
\label{sigtu_visc_small_r}
\eeqs
Hence,
\beqs
\lim_{r \to 0} \sigma^{(t)}_{\rm V} &=&
\lim_{r \to 0} \sigma^{(u)}_{\rm V} = -\lim_{r \to 0} \sigma^{(tu)}_{\rm V} 
 \cr\cr
 &=& \frac{4\pi\sigma_0}{3} \ .
\label{sig_visv_parts_r_0}
\eeqs
This is analogous to the relation that we found for the individual
contributions to $\sigma$ in Eq. (\ref{sigt_sigu_r_0}).  Thus, the
full viscosity cross section has the small-$r$ series expansion 
\beqs
\sigma_{\rm V} &=& \frac{4\pi\sigma_0}{3}\bigg [ 1 - r + r^2 +  O(r^3)
  \bigg ]  \quad {\rm for} \ r \ll 1 \ .
\cr\cr
&&
\label{sigma_visc_nonrel_small_r}
\eeqs
At large $r$, $\sigma_{\rm V}$ has the series expansion 
\beqs
\sigma_{\rm V} &=& \frac{8\pi\sigma_0}{r^2}\bigg [ 2\ln r - 5 +
  \frac{2(3\ln r + 1)}{r} + O\Big ( \frac{\ln r}{r^2} \Big ) \bigg ] 
\cr\cr && {\rm for} \ r \gg 1 \ . 
\label{sigma_visc_large_r}
\eeqs
The prefactor in Eq. (\ref{sigma_visc_large_r}) is 
$8\pi\sigma_0/r^2 = 8\pi \alpha_\chi^2/(\beta_{\rm rel}^4 m_\chi^2)$.

For small $r$, the ratio $\sigma_{\rm V}/\sigma^{(t)}_{\rm V}$
behaves as 
\beq
\frac{\sigma_{\rm V}}{\sigma^{(t)}_{\rm V}} = 1+\frac{r^2}{10} + O(r^3) \ , 
\label{sigvisc_sigtvisc_ratio_small_r}
\eeq
while for $r \gg 1$,
\beq
\frac{\sigma_{\rm V}}{\sigma^{(t)}_{\rm V}} = 2 - \frac{1}{\ln r}
+ O\Big ( \frac{1}{(\ln r)^2} \Big ) \ .
\label{sigvisc_sigtvisc_ratio_large_r}
\eeq

In Fig. \ref{sidm_viscosity_comp_figure} we
plot $\sigma_{\rm V}$ in comparison with $\sigma^{(t)}_{\rm V}$
over the same range of $\beta_{\rm rel}$ as for the regular CM cross
section. A notable feature of these numerical calculations, which is in
agreement with our analytic results, is that for values of $v_{\rm rel} \sim
O(10^3)$ km/sec typical of galaxy clusters, $\sigma_{\rm V}$ is considerably
smaller than $\sigma_{\rm T}$.  This is also evident in Table
\ref{table_COM_comp}. 

%
%
\begin{figure*}[htb!]
	\begin{center}
		\begin{subfigure}{0.46\textwidth}
			\centering
			\includegraphics[width=\textwidth]{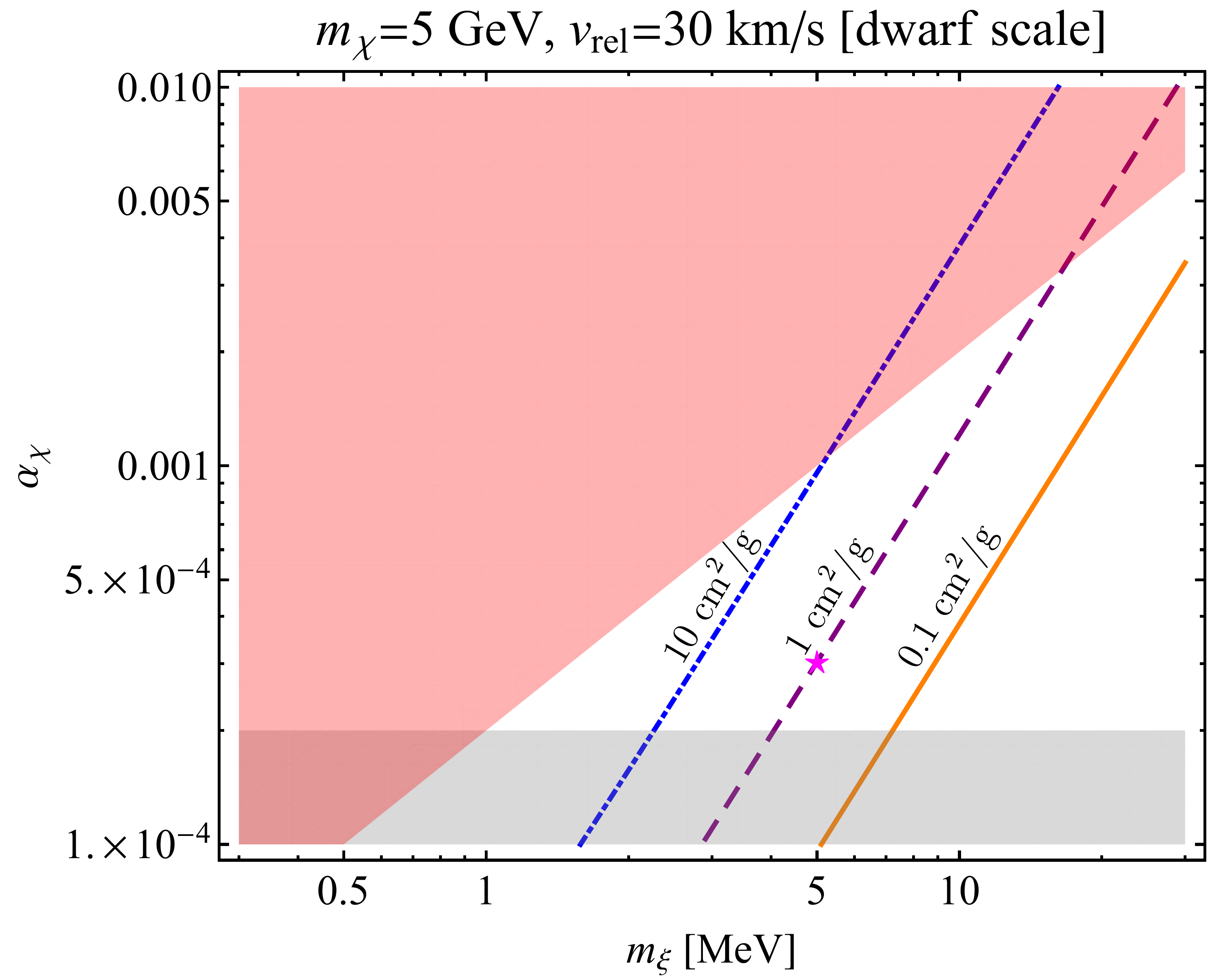}
			\subcaption{$\sigma_T/m_\chi$ contours: dwarf scale}
			\label{sidm_sigmaT_dwarf}
		\end{subfigure}
		\hfill
		\begin{subfigure}{0.46\textwidth}
			\centering
			\includegraphics[width=\textwidth]{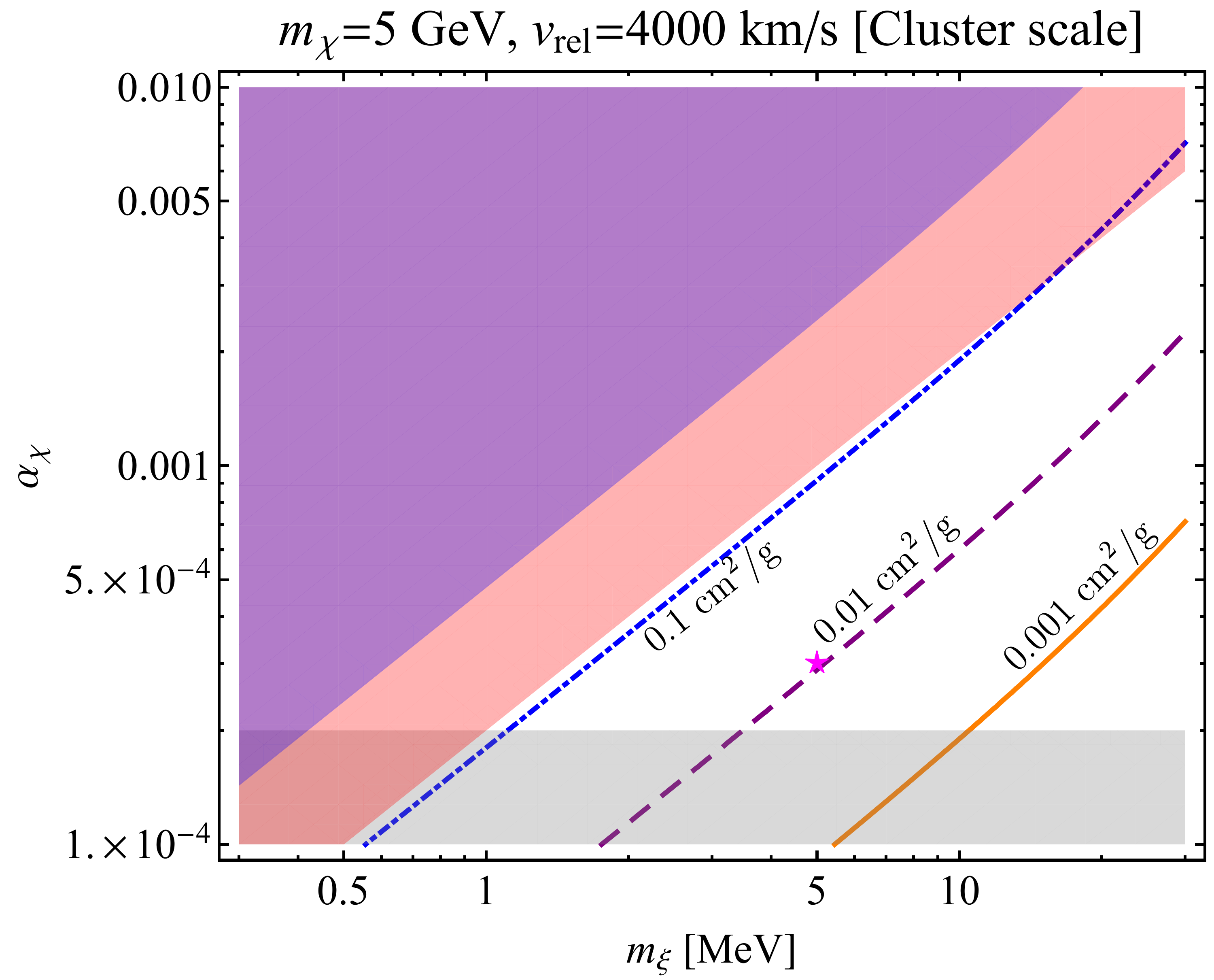}
			\subcaption{$\sigma_T/m_\chi$ contours: cluster scale}
			\label{sidm_sigmaT_cluster}
		\end{subfigure}
		\hfill
		\begin{subfigure}{0.46\textwidth}
			\centering
			\includegraphics[width=\textwidth]{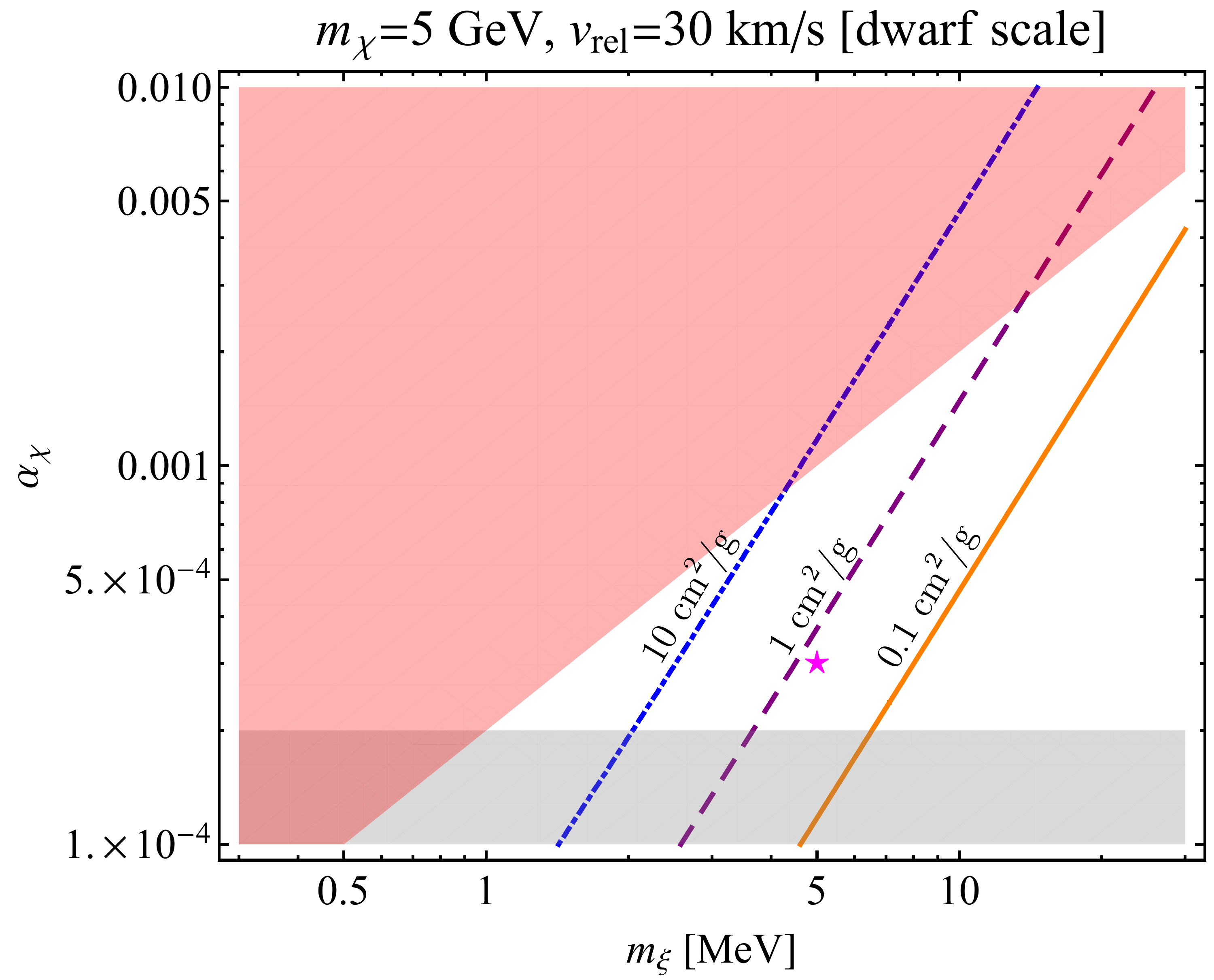}
			\subcaption{$\sigma_V/m_\chi$ contours: dwarf scale}
			\label{sidm_sigmaV_dwarf}
		\end{subfigure}
		\hfill
		\begin{subfigure}{0.46\textwidth}
			\centering
			\includegraphics[width=\textwidth]{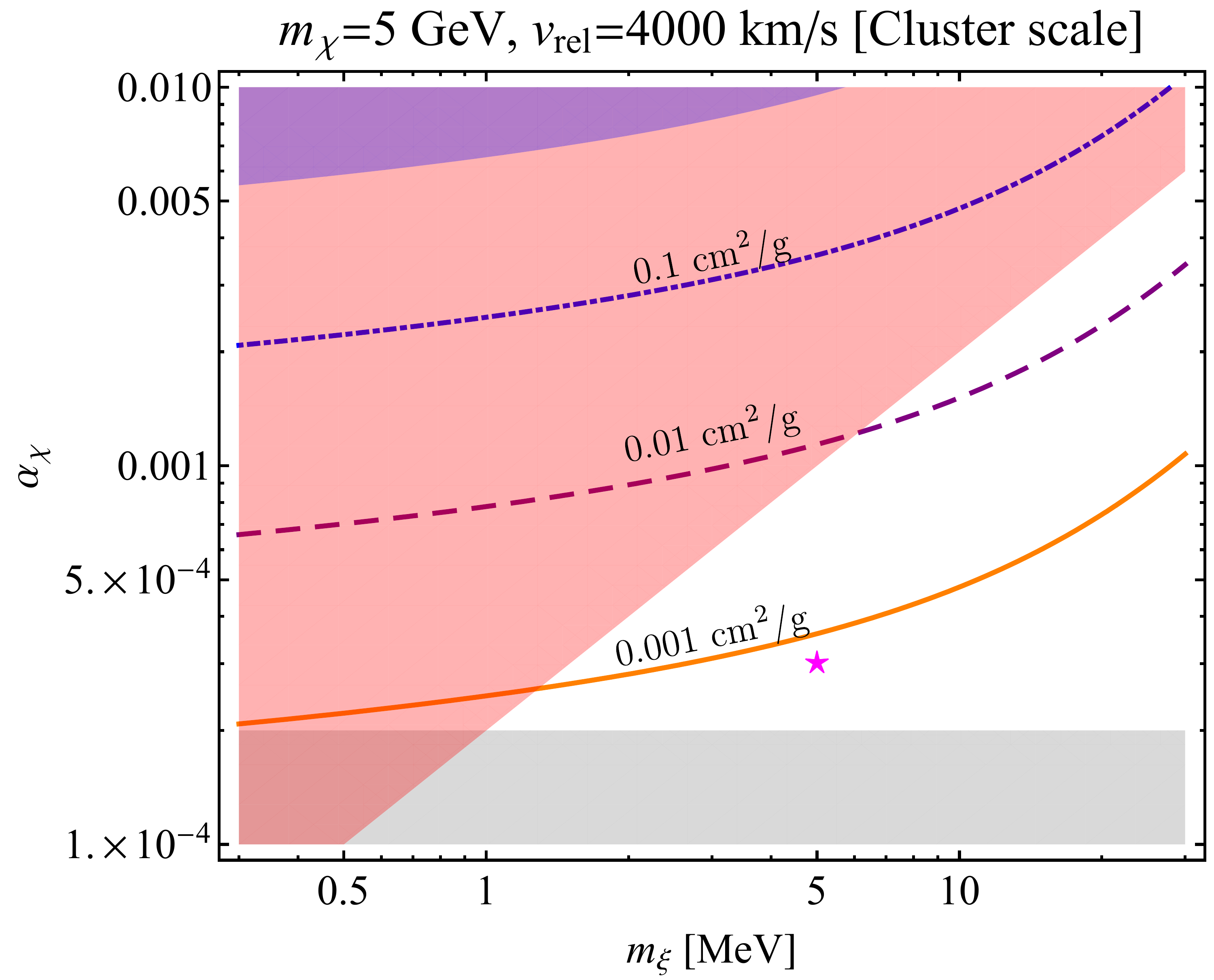}
			\subcaption{$\sigma_V/m_\chi$ contours: cluster scale}
			\label{sidm_sigmaV_cluster}
		\end{subfigure}
	\end{center}
	\caption{\footnotesize Plots showing contours of fixed
          transfer cross section $\sigma_T = \sigma$ and viscosity
          cross section $\sigma_V$, divided by DM mass $m_\chi$, in
          the space of parameters $(m_\xi,\alpha_\chi)$. Our results
          are calculated with the inclusion of both $t$-channel and
          $u$-channel contributions. The left two figures in each
          horizontal row apply for the typical DM-DM relative velocity
          $v_{\rm rel}=30$ km/s in dwarfs, while the right two figures
          apply for the typical velocity $v_{\rm rel}=4 \times 10^3$
          km/s in galaxy clusters. The coupling $\alpha_\chi$ should
          lie above the gray shaded region to satisfy the condition
          $\langle \sigma v \rangle_{\bar\chi \chi \to \xi \xi}
          \gtrsim 0.6 \times 10^{-25}$ cm$^3$/s in order to
          effectively deplete away the symmetric component of the DM
          in the early Universe. The red shaded region is outside the
          Born regime, namely where $\alpha_\chi m_\chi/ m_\xi > 1$,
          and the blue shaded region corresponds to the exclusion
          limit from the Bullet cluster (Galaxy Cluster 1E
          0657–56). The dot-dashed blue
          contour corresponds to $\sigma_T/m_\chi=10$ cm$^2$/g,
          whereas the dashed purple and solid orange contours
          correspond to $\sigma_T/m_\chi=1$ cm$^2$/g and $0.1$
          cm$^2$/g, respectively, and similarly with
          $\sigma_V/m_\chi$.  In each plot, our parameter choice in
          Eq. (\ref{parameter_values}) is indicated by the magenta
          asterisk.}
	\label{sidm_parameter_space}
\end{figure*}
%
%


\section{$\chi \chi \to \chi \chi$ Scattering Cross Sections with
  Vector Mediator}
\label{vector_mediator_section}

In this section we consider the case of a vector mediator with the
SIDM interaction (\ref{chi_chi_v}).  The differential cross section
in this case is just the analogue of the M\"oller cross section with
the photon replaced by the massive vector boson $V$: 
\begin{widetext}
\beqs
\bigg ( \frac{d\sigma}{d\Omega}\bigg )_{\rm CM} &=& \frac{\alpha_\chi^2}{2s}
\bigg [ \frac{s^2+u^2 -4m_\chi^2(s+u-t) + 8m_\chi^4}{(t-m_V^2)^2} + 
        \frac{s^2+t^2 -4m_\chi^2(s+t-u) + 8m_\chi^4}{(u-m_V^2)^2} \cr\cr
        &+&\frac{ 2\{s^2-8m_\chi^2s +12m_\chi^4\} }{(t-m_V^2)(u-m_V^2)}
          \ \bigg ] \ . 
\label{dsigma_domega_vector}
\eeqs
\end{widetext}
In the nonrelativistic limit that is relevant for fitting observational data,
this differential cross section becomes the same as the result for an SIDM
model with a scalar mediator, Eq. (\ref{dsigma_domega_nonrel}),
with the replacement $m_\phi \to m_V$:
\beq
\bigg ( \frac{d\sigma}{d\Omega}\bigg )_{\rm CM,vec} = 
\bigg ( \frac{d\sigma}{d\Omega}\bigg )_{\rm CM,\phi}  \quad {\rm with} \
m_\phi \leftrightarrow m_V \ {\rm for} \ \beta_{\rm rel} \ll 1 \ ,
\label{sigma_scalar_vector_relation}
\eeq
where we append subscripts to indicate vector (vec) versus scalar
mediators.  Quantitatively, the difference between
$(d\sigma/d\Omega)_{\rm CM,vec}$ and $(d\sigma/d\Omega)_{\rm CM,\phi}$
is a term of $O(\beta_{\rm rel}^2)$. Even at the length scale of a few
Mpc in galaxy clusters, $\beta_{\rm rel} \sim 10^{-2}$, and therefore
this difference is negligibly small.  Consequently, our analysis in the
previous section also applies to this model. Similar comments apply
for the transfer and viscosity cross sections.


\section{Study of Parameter Variation}
\label{parameter_variation_section}

In this section we study the dependence of the cross sections divided
by DM mass for reaction (\ref{chichi_reaction}) (calculated with both
the $t$-channel and $u$-channel contributions) on the values of the
coupling, $\alpha_\chi$, and mediator mass, $m_\xi$. In
Fig. \ref{sidm_parameter_space} we show plots of $\sigma/m_\chi =
\sigma_{\rm T}/m_\chi$, and $\sigma_{\rm V}/m_\chi$ as functions of
$\alpha_\chi$ and $m_\xi$. For this study, it will suffice to keep
$m_\chi$ fixed at the value of 5 GeV as in
Eq. (\ref{parameter_values}). The figures in the upper and lower panel
are for $\sigma/m_\chi = \sigma_{\rm T}/m\chi$, and $\sigma_{\rm
  V}/m_\chi$, respectively.  In each horizontal panel, the figures on
the left and right are for the value $v_{\rm rel}=30$ km/s typical of
dwarf satellite galaxies and the value $4 \times 10^3$ km/s typical of
galaxy clusters, respectively.  In each figure we show curves of the
respective cross section divided by $m_\chi$ for the values 10
cm$^2$/g, (dot-dashed blue), 1 cm$^2$/g, (dashed purple), and 0.1
cm$^2$/g, (solid orange). The coupling $\alpha_\chi$ should lie above
the grey region in order to satisfy the bound $\langle \sigma
v\rangle_{\bar\chi\chi \to \xi\xi} \gtrsim 0.6 \times 10^{-25}$
cm$^3$/s from the depletion of the symmetric component constraint on
this ADM model, as discussed in Section \ref{background_section}. The
region shaded red is outside the Born regime and corresponds to
$\alpha_\chi m_\chi/m_\xi > 1$.  The region shaded blue is excluded by
observational data on the Bullet Cluster (Galaxy Cluster 1E 0657-56)
\cite{bullet_cluster_constraints,tulin_yu_review}.  Our parameter
values in Eq. (\ref{parameter_values}) are indicated by the
magenta-colored asterisk.  These plots show how $m_\chi$ and $m_\xi$
can be varied while retaining cross section values that avoid excluded
regions. For a given choice of parameter values, our calculations
(with inclusion of both $t$-channel and $u$-channel contributions)
yield $\sigma_{\rm V} \ll \sigma_{\rm T}$ at $v_{\rm rel}$ values
characteristic of galaxy clusters. In both cases, our resulting cross
sections are in accord with upper limits on $\sigma/m_{\rm DM}$
inferred from fits to properties of galaxy clusters.


\section{Conclusions}
\label{conclusion_section}

In summary, in this paper we have studied a model with
self-interacting dark matter consisting of a Dirac fermion $\chi$
coupled to a scalar or vector mediator such that the reaction $\chi +
\chi \to \chi + \chi$ is well described by perturbation theory.  An
asymmetric dark matter framework is assumed for this study. We have
computed the scattering cross section for this reaction including both
$t$-channel and $u$-channel contributions and have analyzed how the
results with inclusion of contributions from both of these channels
compare with a calculation that has often been used in the literature
that only includes the $t$-channel contribution. Our results elucidate
the interplay between the terms $|{\cal M}^{(t)}|^2$, $|{\cal
  M}^{(u)}|^2$, and the interference term $-2{\rm Re}({\cal
  M}^{(t)*}{\cal M}^{(u)})$ in both the differential and total cross
sections.  We find a particularly strong deviation at large $r$ from
results in the literature for the transfer cross section $\sigma_T$
that include only $t$-channel contributions.  With the illustrative
values of the dark matter fermion mass $m_\chi$, the mediator mass
$m_\xi$, and the coupling $\alpha_\chi$ used here, the region of large
$r$ corresponds to DM velocities $v_{\rm rel} \sim 10^3$ km/s, which
occur in galaxy clusters.  Further, we have studied how our cross
section calculations vary for a range of mediator mass $m_\xi$ and
DM-mediator coupling $\alpha_\chi$. Our analytic and numerical
calculations should be useful in fits to observational data. A
self-interacting dark matter model of the type considered here remains
an appealing approach to accounting for this data on scales ranging
from 1-10 kpc in galaxies to several Mpc in galaxy clusters.


\begin{acknowledgments}

  One of us (R.S.) thanks Prof. S. Nussinov for useful discussions.
  This research was supported in part by the U.S. National Science
  Foundation Grant NSF-PHY-1915093 (R.S.).

\end{acknowledgments}


\begin{appendix}

\section{Condition for the validity of the Born approximation}
\label{born_appendix}

In this appendix we discuss further some aspects of the $\chi + \chi
\to \chi + \chi$ reaction.  We comment first on the relation between
our full quantum field theoretic calculation and the nonrelativistic
quantum mechanical analysis in the nonrelativistic limit, where one
considers scattering of the $\chi$ particle in a potential.  This
relation is relevant since the velocities that occur, both on length
scales of galaxies ($v_{\rm rel} \sim 30-200$ km/s), and on length
scales relevant for galaxy clusters ($v_{\rm rel} \sim O(10^3)$
km/s), are all nonrelativistic.  A standard reduction of a two-body
problem of the scattering of two different particles $a$ and $b$
expresses this in terms of an effective one-body problem in which a
particle with the reduced mass $\mu = m_a m_b/(m_a+m_b)$ 
undergoes a scattering due to an isotropic potential $V$.  For the
equal-mass situation under consideration here, the particle has $\mu =
m_\chi/2$ and velocity $v_{\rm rel}=2v_\chi$, and hence momentum $p =
\mu v_{\rm rel} = (m_\chi/2)(2 v_\chi) = m_\chi v_\chi = |{\vec p}_\chi|$,
where $|{\vec p}_\chi|$ was given in Eq. (\ref{pif}). The corresponding
magnitude of the wavevector is $k = p/\hbar \equiv p$ in our units with
$\hbar=1$.
  
A common approach is to use the Born approximation to describe a
sufficiently weak scattering process.  The condition for the Born
approximation to be valid in the quantum mechanical analysis of
potential scattering takes two different forms depending on $|{\vec
  p}|$.  In both cases, it is essentially the condition that the
scattered wave is a small perturbation on the incident plane wave.  We
use the fact that in this quantum mechanical approach, the interaction
mediated by $\xi$ exchange is represented by a potential,
\beq
V({\vec x}) = V(|{\vec x}|) =
V_0 \frac{e^{-m_\xi |{\vec x}|}}{m_\xi |{\vec x}|}
\label{potential}
\eeq
with
\beq
\frac{V_0}{m_\xi} = \alpha_\chi \ . 
\label{alfrel}
\eeq
We define the distance $|{\vec x}| \equiv d$.  The range of this
potential is $\sim a=1/m_\xi$. The condition for the validity of the Born
approximation takes the following two forms \cite{merzbacher},
depending on the value of $ka = p/m_\xi = \beta_\chi m_\chi/m_\xi =
\sqrt{r}/2$, where $r$ is the ratio (\ref{r}). For $r \ll 1$, the
condition is that the kinetic energy $1/(2\mu a^2)$ should be much
larger than the potential energy $\sim V_0$, i.e., $2 \mu a^2 V_0 \ll
1$.  Substituting $a=1/m_\xi$ and the expression for $V_0$ in
Eq. (\ref{alfrel}), this is the inequality
\beq
\frac{2 \mu V_0}{m_\xi^2} = \frac{\alpha_\chi m_\chi}{m_\xi} \ll 1 \ .  
\label{born1}
\eeq
For  $r \gg 1$, the condition is 
$(V_0 a/\beta_{\rm rel})\ln(2ka) \ll 1$, which can be rewritten as 
\beq
\frac{\alpha_\chi}{\beta_{\rm rel}}
\ln ( \sqrt{r} ) \ll 1 \ .
\label{born2}
\eeq
To show that our parameter choices in Eq. (\ref{parameter_values})
satisfy these conditions, we first consider values of $v_{\rm rel}
\sim 30$ km/s relevant for SIDM dynamics within galaxies.  Then
$\beta_{\rm rel} = 10^{-4}$ so $r=(\beta_{\rm rel} m_\chi/m_\xi)^2 =
10^{-2}$.  Since this is $\ll 1$, condition (\ref{born1}) is
applicable.  We have $\alpha_\chi m_\chi/m_\xi = 0.3$ for our choices
of parameters in Eq. (\ref{parameter_values}). For a value of $v_{\rm
  rel} \sim 3 \times 10^3$ km/s relevant for galaxy clusters,
$\beta_{\rm rel}=10^{-2}$, so $r=10^2$, and hence condition (\ref{born2})
applies. For this value of $v_{\rm rel}$, the left-hand side of the
inequality (\ref{born2}) is 0.069, which
is $\ll 1$.  Thus, as stated in the text, with our choices of
$\alpha_\chi$, $m_\chi$, and $m_\xi$ and for the values of $v_{\rm
  rel}$ of relevance to SIDM effects on scales ranging from 1-10 kpc
in the core of a galaxy to several Mpc for clusters of galaxies, our
restriction to the Born regime is justified.


\section{Quantum Mechanical Treatment of the Yukawa Potential}
\label{yukawa_appendix}

Here we review the quantum mechanical treatment of the Yukawa
potential and derive Eq. (\ref{sigma_t_transfer_nonrel}) for the
transfer cross section from the partial wave analysis. These are
well-known results (e.g., \cite{merzbacher,gw}), but we briefly
discuss them here for the convenience of the reader in comparing the
quantum mechanical treatment with the quantum field theory results. In
a quantum mechanical analysis, one writes the full wave function as
consisting of an incident term (choosing the initial direction of
propagation to be along the $z$ axis, with no loss of generality) plus
the spherical wave due to the scattering by the potential.  For large
distance $d$ from the origin, this has the form
\beq
\psi(\vec{x}) = e^{i k z} + f(\theta) \frac{e^{ikd}}{d} \ ,
\label{psi_eqn}
\eeq
where $k=|{\vec k}|$ is the magnitude of the wave vector of the
  incident particle and we have assumed azimuthal symmetry. The
  scattering amplitude $f(\theta)$ can be expanded in terms of partial
  waves as
\beq
f(\theta) = \frac{1}{k} \sum_{\ell=0}^{\infty}
(2 \ell + 1 )  A_\ell P_\ell (\cos \theta) \ , 
\label{partial_eq}
\eeq
where $P_\ell(\cos\theta)$ is the Legendre polynomial and 
\beq
A_\ell = e^{i \delta_\ell} \sin \delta_\ell \ 
\label{aell}
\eeq
is the quantum mechanical scattering amplitude in the $\ell$'th partial wave,
with phase shift $\delta_\ell$. 
The differential scattering cross section is then 
\begin{align}
  	\nonumber
  	\frac{d\sigma}{d \Omega} &= |f(\theta)|^2 \\
  	& = \frac{1}{k^2} \sum_{\ell, \ell'=0}^\infty
        (2 \ell+1) (2 \ell'+1) A_\ell A_{\ell'}^\ast
        P_\ell (\cos \theta) P_{\ell'} (\cos \theta) \ .
\label{cross_partial_eqn}
\end{align}

Given a potential $V(\vec x')$, the Born approximation to $f$ is
\beq
f = -\frac{\mu}{2\pi} \int d^3 {\vec x'} \, e^{-i {\vec k'} \cdot {\vec x'}}
\, V({\vec x'}) \, e^{i{\vec k} \cdot {\vec x'}}
\label{fborn}
\eeq
where $\vec k$ and $\vec k'$ are the wave vectors of the incident and
scattered waves. This is evidently the Fourier transform of $V(\vec x')$
with respect to the momentum transfer $\vec q = {\vec k} - {\vec k'}$,
with magnitude
\beq
q \equiv |\vec q| = 2k \sin (\theta/2) \ .
\label{q}
\eeq

Consider the Yukawa potential (with $d=|{\vec x}|$):
\beq
V(d) = \pm \alpha_\chi \frac{e^{-m_\xi d}}{d} \ .
\eeq
A standard calculation yields the scattering amplitude
\beq
f_{\rm Yuk} (\theta) =  \mp \frac{2 \mu \alpha_\chi}{m_\xi^2 + q^2} \ .
\label{fyuk}
\eeq
For our application to $\chi$-$\chi$ scattering, the reduced mass is
$\mu = m_\chi/2$ and $k=(m_\chi/2) v_{\rm rel}$, i.e., $q = m_\chi
v_{\rm rel} \sin (\theta/2)$. Therefore, from
Eq. (\ref{cross_partial_eqn}), in the Born approximation, 
\beqs
\bigg(	\frac{d\sigma}{d\Omega}\bigg)_{\rm Yuk} &=&
\frac{\alpha_\chi^2 m_\chi^2}
     {(m_\xi^2+ m_\chi^2 v_{\rm rel}^2 \sin^2 (\theta/2))^2} \cr\cr
     &=& \frac{\sigma_0}{(1+ r \sin^2(\theta/2))^2} \ ,
  	\label{sig_qm_yuk_eqn}
\eeqs
where we have used the definitions of $\sigma_0$ and $r$ in 
Eqs. (\ref{sigma0}) and (\ref{r}). Comparing
Eq. (\ref{sig_qm_yuk_eqn}) with Eq. (\ref{dsigma_domega_nonrel}), one
sees that if one were to approach the calculation without proper use
of the antisymmetrization of the quantum mechanical wave function, then
the Yukawa potential would correspond to inclusion of only the 
$t$-channel contribution to the full quantum field theoretic amplitude.
Finally, applying the definitions of transfer and viscosity
cross sections, given in Eqs. (\ref{transfer_sigma},
\ref{viscosity_sigma}) yields the corresponding cross sections for
this Yukawa potential:
\beq
\sigma_{\rm CM, Yuk} =  \frac{4 \pi \sigma_0}{1+r} \ ,
\label{sigma_yuk}
\eeq
\beq
\sigma_{\rm T, Yuk} = \frac{8 \pi \sigma_0}{r}
\bigg[- \frac{1}{1+r} + \frac{\ln(1+r)}{r} \bigg] \ ,
\label{sigma_transfer_yuk}
\eeq
\beq
\sigma_{\rm V, Yuk} = \frac{16  \pi \sigma_0}{r^2}
\bigg[-2+(2+r)  \frac{\ln(1+r)}{r}   \bigg] \ .
\label{sigma_viscosity_yuk}
\eeq
Thus, these are the cross sections that one would get in a quantum mechanical
treatment if one did not take account of the necessity of antisymmetrizing the
wave function under exchange of identical fermions. 

The calculation in nonrelativistic
quantum mechanics for identical fermion scattering must, of course,
respect the Pauli exclusion principle. In other words, the
wavefunction for the $\chi-\chi$ system should be completely
antisymmetric, i.e., should have the form of a Slater determinant,
namely
\beq
\Psi (x_1, x_2) = \frac{1}{\sqrt{2}} \begin{vmatrix}
	\chi_1 (x_1) & \chi_2 (x_1) \\
	\chi_1 (x_2) & \chi_2 (x_2)
	\end{vmatrix} \ .
\label{slater}
\eeq
From here, it is evident that the normalization factor $1/\sqrt{2}$ in
the Slater determinant wavefunction is equivalent to the factor 1/2 in
the formula for the scattering cross section (\ref{sigma_eq}). The
antisymmetrization in the Slater determinant is the quantum mechanical
equivalent of the inclusion of both the $t$-channel and the
$u$-channel diagrams in the quantum field theoretic calculation. Thus,
a quantum mechanical treatment with proper antisymmetrization for
scattering of identical fermions gives the same result as the
(nonrelativistic limit of the) quantum field theoretic calculation.
We have presented the results for these cross sections for the Born
regime in the text, as Eqs. (\ref{sigma_nonrel}),
(\ref{sigma_sigma_transfer_equality}), and (\ref{sigma_visc_nonrel}).

\end{appendix} 
\bibliographystyle{apsrev4-2}
\bibliography{siadm}
\end{document}